# Streamline-Based Simulation of Carbon Dioxide Sequestration in Saline Aquifers

Feyi Olalotiti-Lawal, Shusei Tanaka and Akhil Datta-Gupta

## Abstract

Subsurface sequestration of CO2 has received attention from the global scientific community in response to climate change concerns due to higher concentrations of CO2 in the atmosphere. Mathematical models have thus been developed to aid the understanding of multiphase flow of CO2 and trapping mechanisms during subsurface sequestration. Solutions to these models have ranged from analytical, semi-analytical and numerical methods, each having its merits and limitations in terms of underlying physics, computational speed and accuracy.

We present a streamline-based method for modeling CO2 transport in saline aquifers which leverages sub-grid resolution capabilities of streamlines in capturing small- and large-scale heterogeneity effects during CO2 injection. Our approach is based on an iterative IMPES scheme and accounts for the physical processes characteristic of CO2 injection in saline aquifers. These include compressibility, gravity, capillarity, mutual solubility, precipitation and formation dry-out effects. Our streamline simulation method provides an extension of previous streamline-based models through rigorous treatment of transverse fluxes arising from compressibility, gravity and capillary effects. We present series of examples encompassing different levels of geologic and geometrical complexity to illustrate the relevance, accuracy and computational efficiency of the approach.

## 1 Introduction

Climate change is a topical environmental issue which has been linked to the rise in global average temperatures because of increase in greenhouse gases (GHG) such as carbon dioxide (CO2) and methane in the atmosphere in the last century. The past few decades have witnessed a global concerted effort at reversing the trend by reducing the concentration of these gases, particularly CO2 in the atmosphere. A promising strategy is Carbon Capture and Storage (CCS) which involves capturing CO2 from industrial sources such as coal fired power plants, transportation of the CO2 and injection in subsurface geologic formations for permanent storage. Candidate formations include depleted hydrocarbon reservoirs, unmineable coal deposits and saline aquifers, which is known to show the highest potential for large scale subsurface storage (USDOE, 2015).

Like most subsurface flow processes, injection of CO2 in saline aquifers poses certain risks (Arts et al., 2008) arising from the dearth of data and subsurface uncertainty. Important risks include leakage of CO2 through old wells or non-sealing faults into underground water or the earth surface. These pose significant hazard to public health and the environment (Apps et al., 2010; Siirila et al., 2012). Proper site selection therefore requires understanding of CO2 plume migration in the short and long terms in the subsurface and understanding flow and trapping mechanisms of CO2 under varying thermodynamic conditions, geologic settings and structures at different time scales. The dominant trapping mechanisms of CO2 including structural, residual, capillary, solubility and mineral have been well documented in the literature (Bachu et al., 2007; Bachu et al., 1994; Ennis-King and Paterson, 2003; MacMinn et al., 2010; Metz et al., 2005; Saadatpoor et al., 2010).

Analytical and numerical predictive models incorporating relevant trapping mechanisms at varying fidelity levels have been developed for reliable evaluation of saline aquifer candidates for CO2 sequestration. Analytical models are valuable in identifying key parameters and/or dimensionless groups for better understanding important physics underlying the process and provides an efficient tool for screening of aquifer candidates for CO2 storage (Mathias et al., 2009b). Nordbotten et al. (2005) presented an analytical solution for CO2 plume evolution during injection in saline aquifers with homogenous media properties and uniform initial conditions. Vilarrasa et al. (2013) proposed a semi-analytical solution as an improvement over the existing analytical solutions by accounting for non-uniform CO2 flux from the well across the aquifer cross-section. Mathias et al. (2009a) and Mijic et al. (2014) included the Forcheimer model in their analytical model to study the influence of non-Darcy flow and gas compressibility on well injectivity. While analytical solutions provide significant benefits in terms of lower computational cost, capturing important details such as small and large scale permeability heterogeneities remains a challenge.

Numerical models allow for more realistic description of subsurface flow of CO2 in saline aquifers. Numerical simulation codes such as TOUGH2_ECO2N (Pruess, 2005), CMG-GEM-GHG (Nghiem et al., 2004) and E300-CO2STORE (Schlumberger, 2014) implicitly/semi-implicitly solve conservation equations using typically finite difference/volume schemes. Therefore, more detailed description of non-linear relationships between model parameters with system state variables such as pressure, temperature and fluid saturations becomes possible. This allows higher fidelity representation of relevant subsurface flow phenomena such as residual and capillary trapping mechanisms (Saadatpoor et al., 2010; Spiteri et al., 2005), gravity induced convection and formation dry out effects (Giorgis et al., 2007; Pruess and Müller, 2009). Capturing these detailed physics, which sometimes require local grid refinements, however often come with much higher computational cost compared with analytical models.

The streamline simulation has been applied (Obi and Blunt, 2006; Qi et al., 2009) to improve upon the computational efficiency of classical finite volume models. Besides faster solutions, streamline models offer visual and physically intuitive representation of the flow, making the results easy to analyze. With streamline-based methods, 3D transport problems are reduced to a set of independent 1D problems oriented along the direction of the total flux (Datta-Gupta and King, 2007). Streamline-based approaches also reduce numerical artifacts and better represent heterogeneities through sub-grid resolutions, making it suitable for large scale and geologically realistic systems, as illustrated in **Fig. 1**. Obi and Blunt (2006) proposed the streamline simulation approach for CO2 sequestration modeling in highly heterogeneous formations, using Henry's Law for CO2 solubility in brine and first order reaction model. Qi et al. (2009) later proposed an improved streamline simulation method to account for mutual solubility of CO2 in aqueous phase and water in CO2-rich phase. Both models showed good agreement with analytical solution in 1D domain, but were not benchmarked with standard simulation models for field scale applications. More importantly, these models have assumed incompressible flow which has been shown to potentially result in erroneous prediction of plume geometry (Vilarrasa et al., 2010).

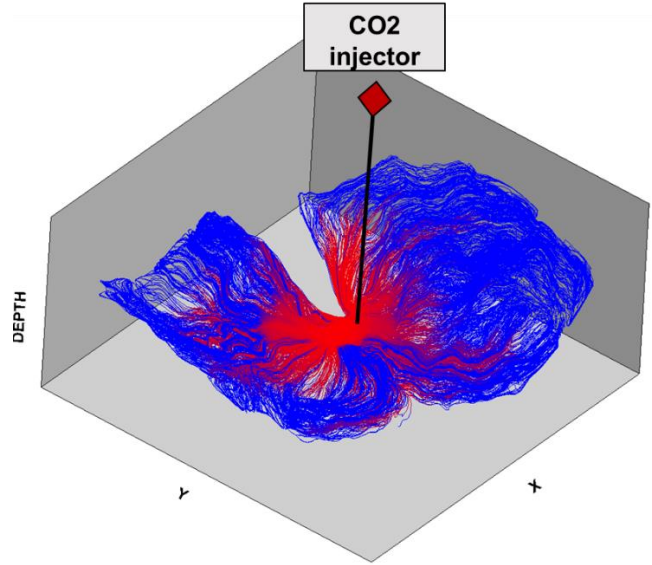

Figure 1: Generated streamline distribution for a CO2 injection in a faulted reservoir. Streamlines are contoured by phase saturations (gas phase in red). This figure illustrates the capability of streamlines in capturing sub-grid resolution

We present a comprehensive streamline-based method for simulation of CO2 sequestration in saline aquifers. Our approach accounts for fluid compressibility effects by incorporating an effective density term that allows for fluid expansion and compression along (Cheng et al., 2006; Osako and Datta-Gupta, 2007). Transverse fluxes such as gravity, capillarity and diffusion are accounted for using the orthogonal projection approach (Tanaka et al., 2014). Mutual solubility effects and precipitation effects are included to model well infectivity alteration during CO2 injection. Comparison between streamline simulation results and commercial compositional finite difference simulation results show good agreement in terms of pressure, phase saturations, component concentrations, while offering improved computational benefit. The outline of this paper is as follows: First we present the simulation model formulation including material balance and fluid property models. Next we provide the steps involved in our streamline simulation approach. We then we present 1D and 2D cross-section illustrative examples. Finally, we demonstrate the robustness of our approach using application to the Johansen field (Eigestad et al., 2009), a candidate for large scale storage of anthropogenic CO2.

## 2 Simulation Model

We begin with the governing equations and transformation into streamline Time-of-Flight (TOF) coordinates (Datta-Gupta and King, 2007). Next we discuss the fluid property and phase equilibria model applied in our approach, followed by modeling of salt precipitation and formation dry-out. The dry-out phenomenon in CO2 sequestration in saline aquifers has been well studied and shown to affect injection and formation pressure (Giorgis et al., 2007; Peysson et al., 2014), which in turn affects CO2 distribution between CO2-rich and aqueous phases – a critical component in the computation of CO2 storage efficiencies. Finally, we discuss the reaction model and formation porosity and permeability variations as a result of salt precipitation and reaction.

### 2.1 Governing Transport Equations

Reactive transport in CO2-brine system during CO2 sequestration in saline aquifers is here modeled using a three-component, two-phase system. Water $(w)$ and CO2 $(c)$ components are distributed in the aqueous $(aq)$ and CO2-rich phases $(g)$ while the salt $(s)$ component is only present in the aqueous phase. Phase and component transport can be modeled by solving a system of coupled nonlinear equations as follows:

$$\frac{\partial}{\partial t}\left(\phi \rho_g s_g\right) + \nabla \bullet \left(\rho_g \mathbf{u}_g\right) = \left(\Delta m_{w,g} - \Delta m_{c,aq}\right) \tag{1}$$

$$\frac{\partial}{\partial t}\left(\phi \rho_{aq} s_{aq}\right) + \nabla \bullet \left(\rho_{aq} \mathbf{u}_{aq}\right) = \left(\Delta m_{c,aq} - \Delta m_{w,g}\right) - \Delta m_{Rxn} \tag{2}$$

$$\frac{\partial}{\partial t}\left(\phi y_w \rho_g s_g\right) + \nabla \bullet \left(y_w \rho_g \mathbf{u}_g\right) = \Delta m_{w,g} \tag{3}$$

$$\frac{\partial}{\partial t}\left(\phi x_c \rho_{aq} s_{aq}\right) + \nabla \bullet \left(x_c \rho_{aq} \mathbf{u}_{aq}\right) = \Delta m_{c,aq} - \Delta m_{Rxn,c} \tag{4}$$

$$\frac{\partial}{\partial t}\left(\phi x_s \rho_{aq} s_{aq}\right) + \nabla \bullet \left(x_s \rho_{aq} \mathbf{u}_{aq}\right) = -\Delta m_{Rxn,s} \tag{5}$$

**Eqs.** 1 and **2** describe phase transport for the CO2-rich $(g)$ and aqueous $(aq)$ phases respectively. **Eqs. 3–5** on the other hand, describe component transport for water dissolved in the CO2-rich phase as well as CO2 and salt dissolved in the aqueous phase respectively. The phase saturations, velocities and densities are denoted by $s_\pi, \mathbf{u}_\pi$ and $\rho_\pi$; $\pi \in \{g, aq\}$ respectively whereas the mass fraction of component $\upsilon$ in aqueous and CO2-rich phases are respectively denoted by $x_\nu$ and $y_\nu$; $\nu \in \{c, s, w\}$. Incremental masses of CO2 dissolved in the aqueous phase and of pure water dissolved in the CO2-rich phase per unit time are respectively denoted as $\Delta m_{c,aq}$ and $\Delta m_{w,g}$. Overall additional mass change in the aqueous phase per unit time as a result of chemical reactions is denoted by $\Delta m_{Rxn}$ while $\Delta m_{Rxn,c}$ and $\Delta m_{Rxn,s}$ denote the additional mass change in aqueous CO2 and salt due to chemical reactions.

Imposing the condition that phase saturations must sum up to unity everywhere at all times, **Eqs. 1** and **2** can be combined to obtain the overall mass balance equation as:

$$\sum_{\pi=g,aq} B_\pi \left[ s_\pi \frac{\partial}{\partial t}\left(\frac{\phi}{B_\pi}\right) + \nabla \bullet \left(\frac{\mathbf{u}_\pi}{B_\pi}\right) \right] = \left(\frac{1}{\rho_{aq} - \rho_g}\right)\left(\Delta m_{c,aq} - \Delta m_{w,g}\right) + \frac{m_{Rxn}}{\rho_{aq}} \qquad (6)$$

Here the phase formation volume factor, denoted by $B_\pi$ is the ratio of phase densities at surface conditions to phase densities at subsurface conditions. Phase velocities can be expressed as function of grid pressure gradient following the usual Darcy's law:

$$\mathbf{u}_\pi = -\lambda_\pi \mathbf{k} \bullet \left(\nabla p_\pi - \rho_\pi g \nabla z\right) \qquad (7)$$

where $\lambda_\pi$ represents the phase mobility, $\mathbf{k}$ the grid permeability tensor and $p_\pi$ the phase pressure which is the difference between system pressure $p$ and gas-water capillary pressure $p_{cgw}$. **Eqs. 6** and 7 can be solved implicitly to obtain instantaneous spatial distribution of the system pressure $p$, which is also taken as the pressure of the non-wetting phase (CO2-rich phase). Note that for this step, $s_\pi, \Delta m_{w,g}, \Delta m_{c,aq}$ and $m_{Rxn}$ values are obtained from the previous time step. This is the implicit step of the IMPES scheme that is applied in our streamline simulation approach. From the pressure solution, the velocity field can be obtained using **Eq. 6**. For the purpose of streamline simulation, the aqueous phase velocity is expressed in terms of total velocity:

$$\mathbf{u}_g = F_g \mathbf{u}_t + \frac{\lambda_{aq}\lambda_g}{\lambda_t} \mathbf{k} \bullet \left(\nabla p_{cg,aq} + \Delta \rho g \nabla z\right) \tag{8}$$

where $F_g$ is the fractional flow of the CO2-rich phase calculated by $\lambda_g/\lambda_t$ while the fractional flow of the aqueous phase is simply obtained as $F_{aq} = 1 - F_g$. Phase velocities can be resolved along the direction of the total velocity by orthogonal projection as a means for effectively accommodating gravity and capillary along streamlines, so that **Eq. 8** becomes (King et al., 2005; Tanaka et al., 2014):

$$\mathbf{u}_g = F_g \mathbf{u}_t + \frac{\lambda_{aq}\lambda_g}{\lambda_t} \mathbf{k} \bullet \left(\nabla p_{cg,aq} + \Delta \rho g \nabla z\right) = f_g \mathbf{u}_t + \mathbf{u}_{g\perp} \tag{9}$$

Where

$$f_g = \frac{\mathbf{u}_t \bullet \mathbf{u}_g}{u_t^2} = F_g + \frac{1}{u_t^2}\frac{\lambda_{aq}\lambda_g}{\lambda_t} \mathbf{u}_t \bullet \mathbf{k} \bullet \left(\nabla p_{cg,aq} + \Delta \rho g \nabla z\right) \tag{10}$$

$$\mathbf{u}_{g\perp} = \left(\mathbf{I} - \hat{\mathbf{u}}_t \hat{\mathbf{u}}_t\right) \bullet \mathbf{u}_g = \frac{\lambda_{aq}\lambda_g}{\lambda_t}\left(\mathbf{I} - \hat{\mathbf{u}}_t \hat{\mathbf{u}}_t\right) \bullet \mathbf{k} \bullet \left(\nabla p_{cg,aq} + \Delta \rho g \nabla z\right)$$

The beauty of the orthogonal projection approach is its robustness and ease of implementation with no requirement of anti-diffusive flux computation as demanded by the operator splitting approach (Berenblyum et al., 2003). Furthermore, for most practical applications, significant portion of the gravity and capillary effects are accounted for along the streamlines, leaving transverse fluxes $\mathbf{u}_{aq\perp}$ of relatively smaller magnitudes to be corrected for on the underlying grid. Larger time step sizes are therefore possible, with overarching effect of improved computational efficiency (Tanaka et al., 2014).

## 2.2 Streamline Simulation

We adopt the generalized representation of a velocity field using a set of bi-streamfunctions and the introduction of the effective density, $\rho$ to account for compressibility effects (Osako and Datta-Gupta, 2007):

$$\rho \mathbf{u}_t = \nabla \psi \times \nabla \chi \tag{11}$$

The introduction of the bi-streamfunctions facilitates an efficient 1D flow description along a new spatial coordinate known as the time of flight, $\tau$. The time of flight is defined as the transit time of a neutral tracer along total velocity $\mathbf{u}_t$, mathematically expressed in the differential form as:

$$\mathbf{u}_t \bullet \nabla \tau = \phi \tag{12}$$

This allows a $(\tau, \psi, \chi)$ coordinate system in which $\mathbf{u}_t$ is perpendicular to both $\nabla \psi$ and $\nabla \chi$. Therefore, it can easily be shown that applying the time of flight definition in **Eq. 12** to a gradient operator in this coordinate system results in the operator: $\mathbf{u}_t \bullet \nabla = \phi \dfrac{\partial}{\partial \tau}$ which can be applied to any scalar field in the model domain. Applying this operator to the flux conservation, knowing that $\nabla \bullet \rho \mathbf{u}_t$ must vanish, we obtain the following form for the velocity divergence which is finite for general compressible systems:

$$\nabla \bullet \mathbf{u}_t = -\phi \frac{\partial \ln \rho}{\partial \tau} = -\sigma \tag{13}$$

The previously discussed transport equations can then be written in the streamline coordinates, applying the orthogonal projection phase fluxes along the direction of the total velocity and the definition of the velocity divergence. The CO2-rich phase transport equation, for instance, becomes:

$$\frac{\partial}{\partial t}\left(\rho_g s_g\right) + \frac{\partial}{\partial \tau}\left(\rho_g f_g\right) = \sigma \frac{\rho_g f_g}{\phi} \tag{14}$$

The second term of the LHS can be rearranged into a form complaint with the 1D solution scheme (Tanaka et al., 2014):

$$\frac{\partial}{\partial \tau}\left(\rho_g f_g\right) = \frac{\partial}{\partial \tau}\left(\rho_g F_g + \frac{\rho_g}{u_t^2}\frac{\lambda_{aq}\lambda_g}{\lambda_t}\left[(\hat{\mathbf{u}}_t \bullet \mathbf{k} \bullet \hat{\mathbf{u}}_t)\left(\phi \frac{\partial p_{cg,aq}}{\partial \tau}\right) + k_z \phi \Delta \rho g \frac{\partial z}{\partial \tau}\right]\right) \tag{15}$$

Where $k_z$ is the z-direction permeability component of the tensor $\mathbf{k}$. Note that compressibility effect turns out to be accounted for along streamlines as an extra sink/source as shown in the right hand side of **Eq. 14**. The CO2-rich phase as well as the component transport equations can be treated similarly. Similarly, component transport equations result in the following 1D predictor equations:

$$\frac{\partial}{\partial t}\left(y_v \rho_g s_g\right) + \frac{\partial}{\partial \tau}\left(y_v \rho_g f_g\right) = \sigma \frac{y_v \rho_g f_g}{\phi} \tag{16}$$

$$\frac{\partial}{\partial t}\left(x_v \rho_{aq} s_{aq}\right) + \frac{\partial}{\partial \tau}\left(x_v \rho_{aq} f_{aq}\right) = \sigma \frac{x_v \rho_{aq} f_{aq}}{\phi} \tag{17}$$

where $v \in \{c, s, w\}$. It is important to point out here that mutual solubility calculations are conducted as phase and saturation fronts are advanced along streamlines. As will be shown later, this is critical to effectively capturing the formation the dry-out phenomena. Phase saturation and component mass fraction profile solutions obtained along the streamlines are mapped on the finite difference grid where transverse fluxes and other physics such as dissolution and reaction are accounted for on the underlying grid using the following corrector forms:

$$\frac{\partial}{\partial t}\left(\phi \rho_g s_g\right) = -\nabla \bullet \left(\rho_g \mathbf{u}_{g\perp}\right) \quad (18)$$

$$\frac{\partial}{\partial t}\left(\phi y_v \rho_g s_g\right) = -\nabla \bullet \left(y_v \rho_g \mathbf{u}_{g\perp}\right) \quad (19)$$

Similar expressions can be written for the aqueous phase and its components. Note that reaction contribution is accounted for here as well on the underlying grid. It is important to note that for practical injection rates of supercritical CO2 into the aquifer, as we have for most CO2 sequestration applications, dominant fluxes are already handled along the streamlines, leaving minimal transverse fluxes (orthogonal to the total flux direction) to be corrected for on the grid. Also in our formulation, only incremental dissolution of components into the phases are included in the equations. This, together with the reaction term in general, represents a small portion of total adjustments on the phase saturation and concentration distributions during the corrector step of our simulation workflow. Consequently, much smaller computational overhead is incurred during the corrector step on the finite difference grid, compared to the predictor step along streamlines.

## 2.3 Multiscale Streamlines

To mitigate grid resolution effects associated with gravity override in subsurface gas injection modeling, we applied a global grid refinement scheme for the transport equations. Similar to previously multiscale streamline simulation methods (Aarnes et al., 2005; Stenerud et al., 2008), we solve for pressure field on a coarse grid and map the flux field to the fine grid for saturation and compositional transport solution. Here, however, the pressure and fields are solved on the original native grid and the coarse flux field is downscaled unto the refined grid for streamline tracing, 1D saturation and composition propagation and traverse flux correction steps. Flux construction on the refined grid followed the Pollock approach using the previously proposed adaptive tracing refinement method (Matringe and Gerritsen, 2004). This was generalized to handle corner point grids by utilizing the usual isoparametric transformations (Datta-Gupta and King, 2007).

In our approach, flow properties in the refine grids, including the flux divergence term, are directly populated with corresponding parent coarse grid values, Also, like the regular Pollock's algorithm, our flux construction remains conservative ($\nabla \cdot \rho \mathbf{q} = 0$), thereby naturally enhancing mass conservation. Finally, at each well completion grid we compute a local pressure distribution $p$ in the refined grid by solving a local pseudo-steady state differential mass balance equation described by:

$$\nabla \cdot \frac{\mathbf{k}}{\mu} \nabla p = c_t \phi V_G \frac{\partial \overline{p}}{\partial t} + q_{well} \qquad (20)$$

Where $q_w$ denote the well completion rate at subsurface conditions, while $\phi$, $\mu$, $c_t$ and $V_G$ respectively represent the porosity, fluid viscosity, total compressibility and bulk volume of the parent coarse well grid at the current time step. Note here that the permeability tensor $\mathbf{k}$ is essentially homogeneous and isotropic. The local pressure distribution is constrained by the average pressure value $\overline{p}$ computed from Eq. 6 on the coarse grid as well as by the coarse cell face fluxes. The solution of **Eq. 20** requires the solution of a single linear system which is only as large as the size of grid refinement. For a three-level vertical grid refinement strategy, **Eq. 20** results only in a series of third order linear systems for each open completion. The solution of all these independent linear systems is inconsequential computationally compared to the global pressure solve. Total fluxes in the refined grid are computed based on the calculated pressure filed using Darcy equation.

The value of our multiscale streamline methodology is illustrated with a scenario in which fluid is being injected at the center of a 2D heterogeneous infinite domain. In **Fig. 2**, we compare the streamlines generated with regular Pollock's algorithm with that generated with our multiscale methodology based on the trajectories and the times of flight along the streamlines. Although equal number of streamlines were traced in both cases, smoother gradation of $\tau$ values can be observed with the multiscale method. With finer $\tau$ discretizations obtained in the multiscale method, we hypothesize a major mitigation of adverse grid resolution effects on achievable streamline sub-grid resolutions during transverse flux corrections (Bratvedt et al., 1996).

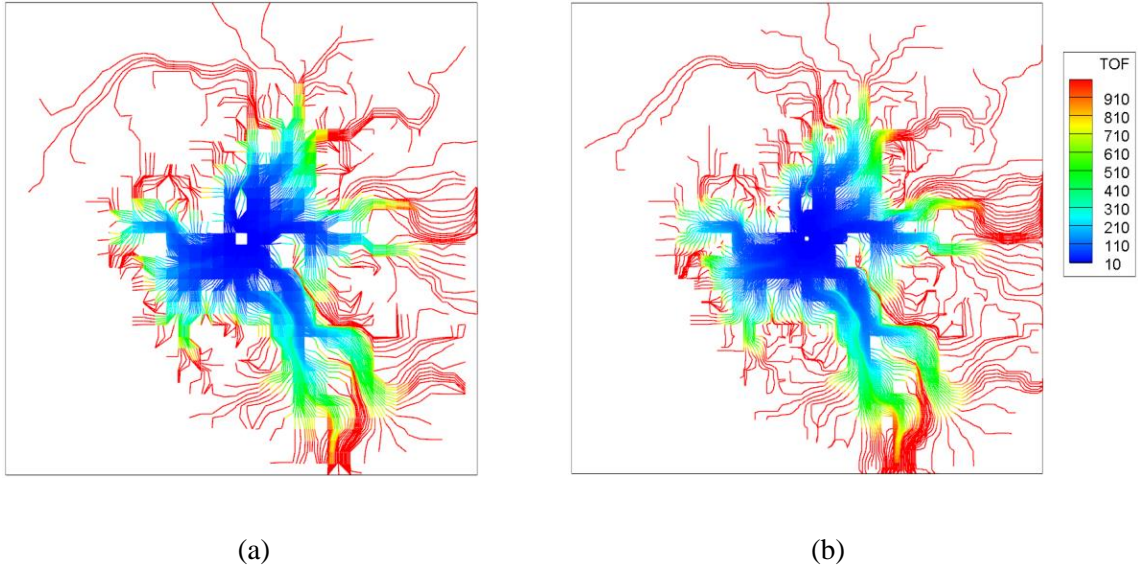

(a)          (b)

Figure 2: Comparing streamline trajectory and time of flight between (a) regular streamline tracing and (b) multiscale streamline tracing

## 2.4 Fluid and Solubility Model

The incremental mass transfer per unit time of CO2 component into the aqueous phase $\Delta m_{c,aq}$ and water into the CO2-rich phase, $\Delta m_{w,g}$ is obtained based on the proposed by Spycher et al. (2003) and Spycher and Pruess (2005) which allows for fast and non-iterative computation of mutual solubility of water and CO2 components in the phases. This extended solubility model which applies for pressure-temperature ranges of interest to CO2 sequestration in saline aquifers, provides equilibrium mole fraction of water dissolved in the CO2-rich phase, $\tilde{y}_w$ and mole fraction of dissolved CO2 in the aqueous phase $\tilde{x}_c$ at specific system pressure, temperature and brine molality using the following equations:.

$$\tilde{y}_w = \frac{K^0_{H_2O} a_{H_2O}}{\Phi_{H_2O} P_{tot}} \exp\left(\frac{(p - p^0)\overline{V}_{H_2O}}{RT}\right) \tag{21}$$

$$\tilde{x}_c = \frac{\Phi_{CO_2}(1 - \tilde{y}_w) P_{tot}}{55.508 \gamma'_x K^0_{CO_2}} \exp\left(-\frac{(p - p^0)\overline{V}_{H_2O}}{RT}\right) \tag{22}$$

where $R$ is the gas constant and $K^0$ denotes thermodynamic equilibrium constants for the distribution of water and CO2 components between aqueous and CO2-rich phases at specified Temperature $T$ and reference pressure $p^0 = 1.0\,bar$, $\Phi$ denotes component fugacity coefficients in the CO2-rich phase, $\overline{V}$ the partial molar volume of pure phases between reference pressure $p^0$ and $p$. The activity of liquid water due to the presence of dissolved salts is denoted by $a_{H_2O}$, while $\gamma'_x$ denotes the activity coefficient of CO2 dissolved in the aqueous phase. This approach assumes infinite dilution of H2O component in the CO2-rich phase for calculating fugacity coefficients using a calibrated Redlich-Kwong (Redlich and Kwong, 1949) equation of state (Spycher et al., 2003). The mass fraction of H2O tends to be ideally very small in the CO2-rich phase makes the assumption reasonable from the mutual solubility computational accuracy standpoint. The computational advantage of the simplifying assumption is that the equilibrium model reduces to a set algebraic equations used for solving mutual solubilities without iteration once system pressure, temperature and salt molality are known. Computed equilibrium mole fractions are converted to mass fractions and in turn, used in the computation of mutual mass solubilities per time between the phases as will be shown in a later section.

Standard phase property correlations were incorporated in our simulation model based on the high accuracy achievable as well as their limits of validity which subsume the ranges of formation pressures and temperatures of interest for CO2 sequestration in saline aquifers Aqueous phase density was obtained as a volume weighted average of brine density $\rho_{Br}(x_{s,Br}, T)$ and dissolved aqueous CO2 density $\rho_{cd}(T)$:

$$\rho_{aq} = \rho_{Br}(x_{s,Br}, T) \left\{ 1 - x_c \left( 1 - \frac{\rho_{Br}(x_{s,Br}, T)}{\rho_{cd}(T)} \right) \right\}^{-1} \quad ; \quad x_{s,Br} = \frac{x_s}{1 - x_c} \tag{23}$$

Where brine density $\rho_{Br}(x_{s,Br}, T)$ is obtained from Dittman (1977) correlation which approximates brine density as a function of brine concentration and system temperature. Aqueous CO2 density, on the other hand is computed from the correlation proposed by Garcia (2001) which provides the corrected molar volume of CO2 due to dissolution in the aqueous phase as a function of Temperature in degree Celsius. The corrected density can then be calculated as a ratio of CO2 molecular weight $M_{CO_2}$ and corrected molar volume:

$$\rho_{cd} = \frac{M_{CO_2}}{37.51 - 0.09585T + 8.740 \times 10^{-4} T^2 - 5.044 \times 10^{-7} T^3} \times 10^6 \tag{24}$$

Aqueous phase viscosity $\mu_{aq}$ is computed from the standard correlation by (Kestin et al., 1981) which gives the dynamic viscosity of aqueous solution of NaCl within the temperature range $20-150°C$ and pressure range $0.1-35 MPa$ with $\pm 0.5\%$ accuracy. The CO2-rich phase density, $\rho_g$ follows the Span and Wagner (1996) model which, based on an equation of state, provides $\rho_g$ at temperatures and pressures up to $1100K$ and $800 MPa$ respectively with $\pm 0.05\%$ accuracy. Dynamic viscosity of CO2-rich phase, $\mu_g$ is computed from the Fenghour et al. (1998) correlation which is valid at temperatures and pressures up to $1000K$ and $300 MPa$ respectively with approximate accuracy of $\pm 0.3\%$. For simplicity, we have assumed negligible effect of dissolved CO2 on $\mu_{aq}$ and of vaporized water on $\mu_g$.

## 3 Simulation Steps

Our streamline-based simulation approach follows an iterative IMPES scheme as shown in **Fig. 3**. At first global iteration, pressure and fluxes are computed based on phase saturations and component concentrations at old time step. New time level phase saturations are obtained from streamline 1D transport solutions followed by corrector solutions as well as equilibrium and reactions calculations on the finite difference grid. The pressure and flux are computed with updated properties in the second global iteration and the process until global residual is less than specified values. The steps involved are discussed in more detail as follows:

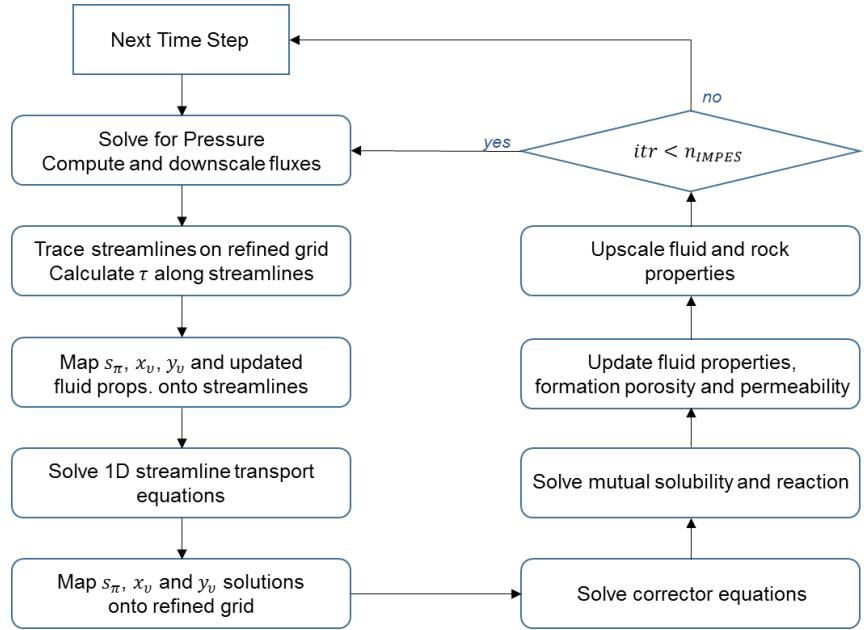

Figure 3: Iterative IMPES scheme for streamline simulation

(1) Solve the overall mass balance equation in **Eq. 6** to obtain the spatial distribution of the field system pressure at the new time level, $p^{n+1}$ using phase saturations and mass fractions at old time level, $s_\pi^n$, $x^n$, $y^n$. Using the pressure solution, compute phase fluxes using **Eq. 8**. Downscale phase fluxes unto the refined grid.

(2) Trace streamlines along total field velocity, obtained from the sum of phase fluxes based on the algorithm proposed by Pollock (1988).

(3) Solve phase and component transport equations along streamlines using **Eqs. 14 –17**. This step results in intermediate phase saturations and component mass fractions $s_\pi^{n+1/2}$, $x^{n+1/2}$, $y^{n+1/2}$ in each streamline segment.

(4) Next is to solve for mutual solubility to redistribute the components between the CO2-rich and aqueous phases by readjusting their respective mass fractions $x^{n+1/2}$ and $y^{n+1/2}$ along each streamline. The following are the steps involved for an arbitrary streamline segment:

   i. Similar to Qi et al. (2009) we start by computing overall initial masses of CO2 $(M_c)$, water $(M_w)$ and salt $(M_s)$ present per bulk volume:

$$M_c = \phi^n \left[ x_c^{n+1/2} \rho_w^{n+1/2} s_w^{n+1/2} + \left(1 - y_w^{n+1/2}\right) \rho_g^{n+1/2} s_g^{n+1/2} \right]$$

$$M_w = \phi^n \left[ \left(1 - x_c^{n+1/2} - x_s^{n+1/2}\right) \rho_{aq}^{n+1/2} s_{aq}^{n+1/2} + y_w^{n+1/2} \rho_g^{n+1/2} s_g^{n+1/2} \right] \tag{25}$$

$$M_s = \left(\phi_0^n - \phi^n\right)\rho_s + \phi^n x_s^{n+1/2} \rho_{aq}^{n+1/2} s_{aq}^{n+1/2}$$

   Where $\rho_s$ denotes salt density and $\phi_0^n$ porosity prior to salt precipitation

   ii. Next is to calculate salt molality, $n_s$ for the equilibrium component mole fractions:

$$n_s = \frac{x_{s,Br}}{1 - x_{s,Br}} \frac{1000}{MW_{salt}}; \quad x_{s,Br} = \frac{x_s^{n+1/2}}{1 - x_c^{n+1/2}} \tag{26}$$

   where $x_{s,Br}$, the salt mass fraction in the aqueous phase without dissolved CO2 component.

   iii. Compute equilibrium mole fractions of CO2 ($\tilde{x}_c$) and H2O ($\tilde{y}_w$) in the aqueous and CO2-rich phases using **Eqs. 21** and **22**. These are then converted to mass fractions:

$$y_w^{n+1} = \frac{\tilde{y}_w . MW_{H_2O}}{\tilde{y}_w . MW_{H_2O} + (1 - \tilde{y}_w) . MW_{CO_2}} \tag{27}$$

$$x_c^{n+1} = \frac{\tilde{x}_c . MW_{CO_2}}{\tilde{x}_c . MW_{CO_2} + \tilde{x}_s . MW_{salt} + (1 - \tilde{x}_s - \tilde{x}_s) . MW_{H_2O}} \tag{28}$$

   iv. Update phase saturations based on the new equilibration mass fractions:

$$s_{aq}^{n+1} = \frac{1}{\phi^n \rho_{aq}^{n+1}} \left[ \frac{M_w - y_w^{n+1}(M_w + M_c)}{(1 - x_s^{n+1/2})(1 - y_w^{n+1}) - x_c^{n+1}} \right]; \quad s_g^{n+1} = 1.0 - s_{aq}^{n+1} \tag{29}$$

Negative phase saturation in this step signifies phase disappearance which necessitates an update in component mass fractions in the remaining phase:

$$\text{If } s_{aq}^{n+1} > 1.0: \ s_g^{n+1} = 0; \ y_w^{n+1} = 0; \ x_c^{n+1} = \frac{M_c}{\phi^n \rho_w}; \ x_s^{n+1/2} = x_{s,Brine}(1 - x_c^{n+1/2}) \tag{30}$$

$$\text{If } s_{aq}^{n+1} < 0.0: \ s_g^{n+1} = 1.0; \ y_w^{n+1} = 1 - \frac{M_c}{\phi^n \rho_g}; \ x_c^{n+1} = 0; \ x_s^{n+1/2} = 0 \tag{31}$$

v. The final step in the mutual solubility calculations is to cater for salt dissolution and precipitation which contributes to the formation dry-out phenomenon. Salt precipitates if salt mass fraction in CO2-free aqueous solution, $x_{s,Br} > x_{s,Br}^{CR}$. Here $x_{s,Br}^{CR}$ denotes the critical salt mass fraction which represents the maximum salt concentration that can dissolve in CO2-free aqueous solution at the system temperature. This is computed from the Potter et al. (1977) correlation which provides $x_{s,Br}^{CR}$ as a function of temperature $T$ in $^\circ C$:

$$x_{s,Br}^{CR} = 0.26218 + 7.2 \times 10^{-5} T[^\circ C] + 1.06 \times 10^{-6} (T[^\circ C])^2 \tag{32}$$

vi. To account for salt precipitation, first total mas of water dissolved in the CO2-rich phase, $m_{wg}$ and of salt dissolved in the aqueous phase, $m_{sw}$ are calculated:

$$m_{wg} = \phi^n \rho_g s_g^{n+1} y_w^{n+1}; \quad m_{sw} = M_s - (\phi_0^n - \phi^n)\rho_s \tag{33}$$

Then used in the computation of $x_{s,Br}$ as follows:

$$x_{s,Br} = \frac{X_s}{1 + X_s}; \quad X_s = \frac{m_{sw}}{M_w - m_{wg}} \tag{34}$$

These are then supplied to the following to algorithm from which the final salt mass fraction update $x_s^{n+1}$, intermediate porosity (due to precipitation) $\phi^{n+1/2}$ and updated salt molality $n_s$ of the aqueous phase are obtained:

If $x_{s,Br} > x_{s,Br}^{CR}$: Salt precipitates out of aqueous phase

$$m_{s,PPT} = \left(\phi^n s_{aq}^{n+1} \rho_{aq} - m_{wg}\right)\left(1 - x_c^{n+1}\right)\left(x_{s,Br} - x_{s,Br}^{CR}\right) \tag{35}$$

$$\phi^{n+1/2} = \phi^n - m_{s,PPT}/\rho_s \; ; \; x_s^{n+1} = \frac{x_{s,Br}^{CR}}{1 - x_c^{n+1}} \; ; \; n_s = \frac{x_{s,Br}^{CR}}{1 - x_{s,Br}^{CR}} \frac{1000}{MW_{salt}} \tag{36}$$

Else If $\phi^n \leq \phi_0^n$: Apparent salt concentration is subcritical and precipitated salt is present

$$m_{s,PPT} = \left(\phi_0^n - \phi^n\right)\rho_s - \left(\phi^n s_{aq}^{n+1} \rho_{aq} - m_{wg}\right)\left(1 - x_c^{n+1}\right)\left(x_{s,Br}^{CR} - x_{s,Br}\right) \tag{37}$$

If $m_{s,PPT} \geq 0$: Part of precipitated salt dissolves, leaving behind $m_{s,PPT}$

$$\phi^{n+1/2} = \phi_0^n - m_{s,PPT}/\rho_s \; ; \; x_s^{n+1} = \frac{x_{s,Br}^{CR}}{1 - x_c^{n+1}} \; ; \; n_s = \frac{x_{s,Br}^{CR}}{1 - x_{s,Br}^{CR}} \frac{1000}{MW_{salt}} \tag{38}$$

Else: All precipitated salt dissolve in aqueous phase

$$\phi^{n+1/2} = \phi_0^n \; ; \; n_s = \frac{M_s}{M_w - m_{wg}} \frac{1000}{MW_{salt}} \tag{39}$$

$$x_s^{n+1} = \frac{\left(1 - x_c^{n+1}\right) n_s \, MW_{salt}/1000}{1 + n_s \, MW_{salt}/1000}$$

End

End

The impact of formation dry out is significant near the CO2 injector. When large volume of dry CO2 comes in contact with the aqueous phase resulting in more water from the aqueous phase vaporizing into the CO2-rich phase. This results in significant increase in $m_{wg}$ and in turn, in $x_{s,Br}$ until a certain point where calculated $x_{s,Br} > 1$ leading to a negative calculated salt molality, $n_s$. At this stage the streamline segment cell completely dries up and phase saturations and component mass fractions are updated as follows:

$$\phi^{n+1/2} = \phi^n \left(1 - \frac{\rho_{aq} s_{aq}^{n+1} x_s^{n+1}}{\rho_s}\right) \tag{40}$$

$$s_{aq}^{n+1} = 0 \,;\, s_g^{n+1} = 1 \tag{41}$$

$$y_w^{n+1} = 1 - \frac{M_c}{\phi^n \rho_g} \,;\, x_c^{n+1} = x_s^{n+1} = 0 \tag{42}$$

(5) Map updated phase saturations, component mass fractions and porosity from streamlines to the refined grid to correct for gravity and capillary effects. Repeat step 4 on the grid. Note that corrections on the grid due to transverse fluxes as well as mutual solubility, and consequently resulting changes in saturations and mass fractions, are very minimal and require a very small fraction of the overall computation cost.

(6) Incremental mass of CO2 dissolved in the aqueous phase $\Delta m_{c,aq}^{n+1}$ and of water dissolved in the CO2-rich phase $\Delta m_{w,g}^{n+1}$ needed for the system pressure update are thus computed as follows using updated phase saturations and component mass fractions.

$$\Delta m_{c,aq}^{n+1} = \phi^{n+1/2} s_w^{n+1} x_c^{n+1} - \phi^n s_w^n x_c^n \tag{43}$$

$$\Delta m_{w,g}^{n+1} = \phi^{n+1/2} s_g^{n+1} y_w^{n+1} - \phi^n s_g^n y_w^n$$

(7) The final update of the grid porosity is obtained after accounting for reaction. Similar to Obi and Blunt (2006) we have assumed a simple first order reaction between aqueous CO2 and dissolved salt to permanently deposit a secondary mineral on the pore walls of the grid. We start by calculating the limiting reaction concentration as follows:

$$C_0 = \rho_{aq} \min\left(x_c^{n+1}/MW_{CO_2}, x_s^{n+1}/MW_{salt}, (1 - x_c^{n+1} - x_s^{n+1})/MW_{H_2O}\right) \tag{44}$$

Therefore changes due to chemical reaction of rate constant $k$ over time step $\Delta t$ can be obtained as:

$$\frac{\phi^{n+1}}{\phi^{n+1/2}} = \frac{\phi_0^{n+1}}{\phi_0^n} = 1 - C_0 s_{aq}^{n+1} MW_{RockSalt} \left(1 - e^{-k\Delta t}\right) / \rho_{RockSalt} \tag{45}$$

Here $MW_{RockSalt}$ and $\rho_{RockSalt}$ denote molecular weight and density of the reaction product – deposited secondary mineral. Note that we have assumed similar percentage change in $\phi$ and $\phi_0$ which are grid porosity with and without aqueous salt precipitation respectively. Incremental aqueous phase and component mass change due to chemical reaction needed for the pressure update is calculated as follows:

$$\Delta m_{Rxn} = C_0 s_{aq}^{n+1} \left(1 - e^{-k\Delta t}\right)\left(MW_{CO_2} + MW_{salt} + MW_{H_2O}\right) \tag{46}$$

$$\Delta m_{Rxn,c} = C_0 s_{aq}^{n+1} MW_{CO_2} \left(1 - e^{-k\Delta t}\right); \quad \Delta m_{Rxn,s} = C_0 s_{aq}^{n+1} MW_{salt} \left(1 - e^{-k\Delta t}\right)$$

(8) Grid permeability is updated as a consequence grid porosity changes due to aqueous salt precipitation and chemical reaction. In our approach we adopted the 'tube-in-series' model proposed by Verma and Pruess (1988). The two-parameter model is based on the fractional length of pore bodies $\Gamma$ and the fraction of original grid porosity $\phi_r$ for which permeability vanishes. Using this model, the fractional change in permeability is given as:

$$\frac{k}{k_0} = \theta^2 \frac{1 - \Gamma + \Gamma/\omega^2}{1 - \Gamma + \Gamma[\theta/\theta + \omega - 1]^2} \tag{47}$$

$$\theta = \frac{1 - S_s - \phi_r}{1 - \phi_r}; \quad \omega = 1 + \frac{1/\Gamma}{1/\phi_r - 1}$$

Where $k_0$ and $k$ are original and updated grid permeability values respectively. For all the examples presented in this paper we have assumed 0.8 for values of $\phi_r$ and $\Gamma$, while solid phase saturation is computed as $S_s = 1 - \phi^{n+1}/\phi_{00}$, where $\phi_{00}$ represent original grid porosity.

(9) Upscale grid properties on the coarse grid for the next global pressure solve. First upscale updated grid porosities using bulk volume weighted averages. Then saturations and concentrations based on pore volume weighted averages. Finally, update coarse scale permeability values based on as a geometric mean of harmonic-arithmetic and arithmetic-harmonic resultant permeability values.

The iterative IMPES approach uses updated phase saturations and component mass fractions through the global iteration process. This leads to better accommodation of the effects of buoyancy and rock property changes due to precipitation and reaction and improves the solution after 2 or 3 iterations based on our observation. Processes with small viscous to gravity ratios will likely require more iterations to reach the specified global convergence criteria. It is worthwhile to note that an alternative algorithm may involve carrying out mutual solubility calculations on finite difference grid after step 5 (Qi et al, 2009). However, as we will show later on, this results in critical phenomena such as salt precipitation and effect on well injectivity being poorly captured especially for strongly water-wet formations.

Once the injection well is shut-in well fluxes go to zero, meaning streamline-based convection calculations are no longer available. As a result, our formulation identically reduces to a conventional finite difference method to account for transverse fluxes (gravity, capillarity and diffusion) and chemical reaction. Simulation results from our streamline-based approach have been compared with that from a commercial finite difference simulator with emphasis on simulation responses during the injection period. This is because accurate prediction of fluid phase and component concentration distribution at injection shut-in is fundamental to reliable modelling of long term CO2 trapping mechanisms post-injection.

## 4  Case Examples

We present cases to validate our streamline-based simulation approach with the CO2STORE module in ECLIPSE reservoir simulator (Schlumberger, 2014). CO2STORE is a dedicated commercial simulator for carbon sequestration in saline aquifers which models all trapping mechanisms and accounts for all relevant physics. We begin with a simple 1D problem and then to 2D homogeneous and heterogeneous cross-section models to demonstrate buoyancy effects on CO2 plume propagation. We have assumed negligible capillarity and molecular diffusion effects for the time being. We have applied the parameters in **Table 1** to all simulation cases presented in this paper. We will discuss the computational benefit afforded by our streamline simulation-based simulation approach compared with finite difference models. Next, we compare the results from mutual solubility calculations on streamline segments and on finite difference grids after the mapping step. We then compare the differences in responses with and without capillarity to illustrate the impact of the using wither solubility calculations. Finally to demonstrate the robustness of our approach, we apply the streamline-based simulation approach to evaluate the Johansen field CO2 sequestration project.

## 4.1 Example 1: 1D Case

The first example models a 1D convection problem. The main essence of this case is to provide a conceptual model for quick illustration of physics associated with CO2 injection while we ignore influence of gravity segregation for the time being, and also to validate the correct handling of phase saturation and component concentration calculations in our method as compared with results from CO2STORE using an equivalent input set. The model comprises a homogeneous rectangular rock piece of length $200\,ft$, discretized into $200$ grid cells of equal dimension $1\,ft \times 1\,ft \times 1\,ft$ as shown in **Fig. 4**. The incompressible rock is assumed to have uniform initial permeability and porosity values of $100\,mD$ and $0.2$ respectively with negligible capillary effects. Initial pressure and brine saturation were $1500\,psia$ and $1.0$ respectively. Dry CO2 is injected at $5.0\,Mscf/D$ for $1.5\,Days$ (equivalent of $0.425\,PVI$ cumulative) while the other end is open to flow and thus, maintained at the initial pressure of $1500\,psia$ to simulate a continuous aquifer and to prevent unwanted premature pressure buildup in the formation.

Table 1: General simulation parameters

| Parameter | Value |
|---|---|
| Aquifer Temperature [$°F$] | 140 |
| Brine Salinity [mass fraction NaCl] | 0.2 |
| First Order Reaction Rate [$/yr$] | 5.0E-04 |
| Salt Density, $\rho_s$ [$lb/ft^3$] | 135 |
| CO2 Surface Density, $\rho_g^\circ$ [$lb/ft^3$] | 0.116 |
| Pure Water Surface density, $\rho_w^\circ$ [$lb/ft^3$] | 62.37 |
| Rock Salt Density, $\rho_{Rocksalt}$ [$lb/ft^3$] | 135 |
| Molecular Weight of Salt, $MW_{salt}$ [$lb/lb-mol$] | 58.4 |
| Molecular Weight of CO2, $MW_{CO_2}$ [$lb/lb-mol$] | 44.0 |
| Molecular Weight of Water, $MW_{H_2O}$ [$lb/lb-mol$] | 18.0 |
| Molecular Weight of Rock Salt, $MW_{Rocksalt}$ [$lb/lb-mol$] | 100 |
| Aqueous Phase Endpoint Relative Permeability, $k_{r,aq}^e$ [-] | 0.95 |
| Aqueous Phase Relative Permeability Exponent, $n_{aq}$ [-] | 4.0 |
| Aqueous Phase Critical Saturation, $s_{wc}$ [-] | 0.1 |
| CO2-rich Phase Endpoint Relative Permeability, $k_{rg}^e$ [-] | 0.4 |
| CO2-rich Phase Relative Permeability Exponent, $n_g$ [-] | 2.0 |
| CO2-rich Phase Residual Saturation, $s_{gr}$ [-] | 0.2 |

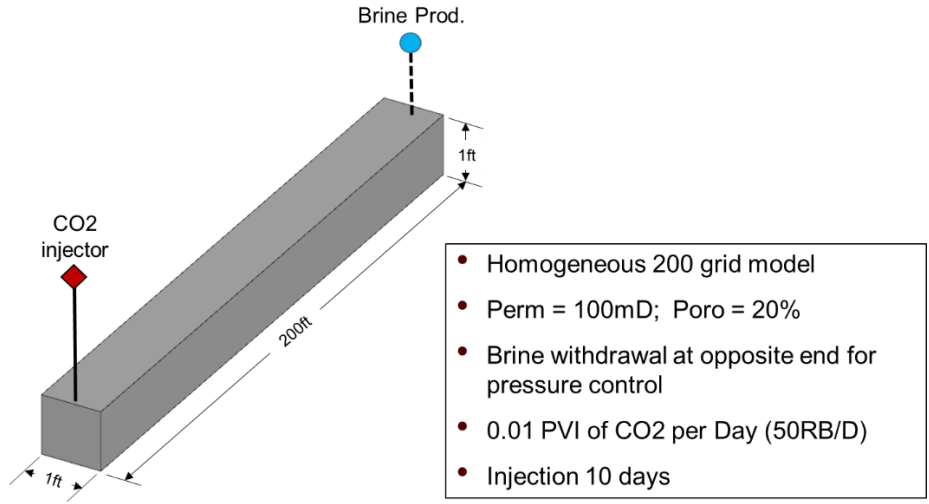

Figure 4: Set-up for the 1D simulation case showing the CO2 injector at one end of the porous of rock and the brine producer well (used to mimic a semi-infinite medium) on the other end

Simulation results in terms of grid pressure, CO2-rich phase saturation and aqueous CO2 and salt mass fractions are compared with CO2STORE at $0.5 Day$, $1.0 Day$ and $1.5 Days$ as shown in **Fig. 5**. All comparisons show excellent agreement with the commercial finite difference simulator. It is easy to notice the effect of fluid compressibility in the pressure profile as indicated by sharp differences in pressure gradients between the two phase region and the brine phase region. Consequently, ignoring compressibility effects of rock and fluid, as in previous streamline-based simulation methods, may result in erroneous flux computations computation leading to incorrect CO2 plume migration predictions.

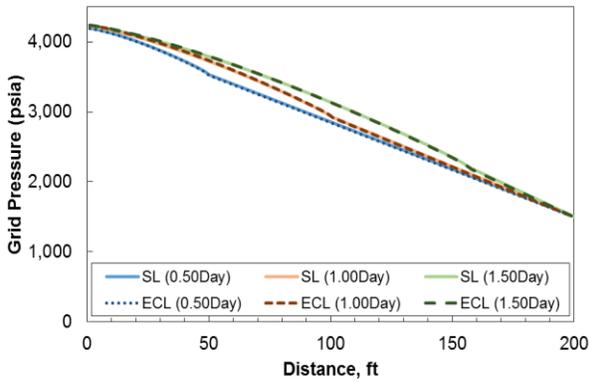
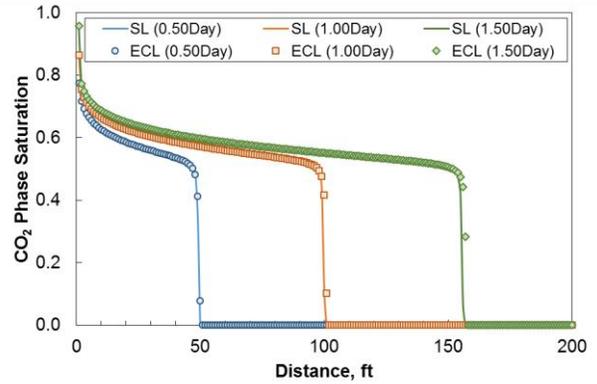

(a)  (b)

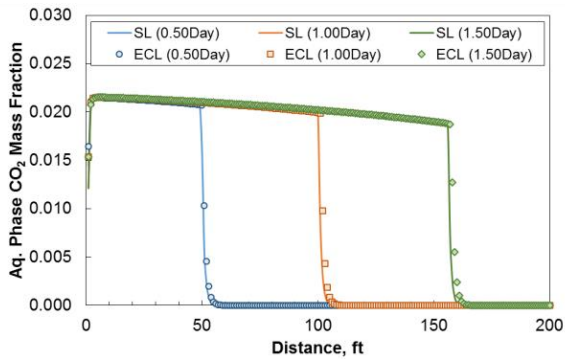
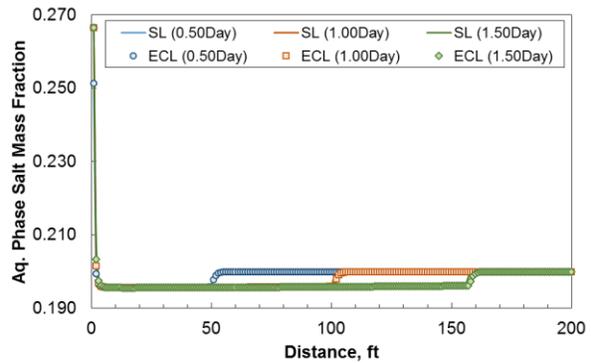

(c)  (d)

Figure 5: Model validation in 1D with commercial FD compositional simulator showing (a) Grid pressure profile (b) CO2-rich phase saturation (c) Aqueous phase CO2 mass fraction, and (d) Aqueous phase salt mass fraction

Behind the two phase front, aqueous phase CO2 concentration shows an expected slight variation with grid pressure the consistent with experimental data (Spycher and Pruess, 2005). An equally important feature well captured by out model is the slow evaporation of water from the aqueous phase into the CO2-rich phase which results in gradual drying out of the grid cells in the vicinity of the CO2 injection well or completion. This causes a sharp increase in the salt concentration and in turn, a decrease in aqueous phase CO2 concentration. The aqueous phase saturation gradually decreases below critical value and, depending on the injection rate and duration, completely vanishes eventually leaving a unit CO2-rich phase saturation.

In streamline based simulation methods, these phenomena are difficult, if not impossible to capture without mutual solubility calculations on streamline segments prior to mapping to the finite difference grid.

## 4.2 Example 2: Homogeneous 2D Cross Section

Nest we demonstrate the developed model with a cross-section case to illustrate the effect of buoyancy during CO2 injection and post-injection periods. Consider a slice of porous media of length $5298\,ft$, width $5.298\,ft$ and depth $50\,ft$ uniformly discretized into $100 \times 1 \times 50$ grids as shown in **Fig. 6(a)**. Here as well, we assume homogeneous permeability and porosity value of $100\,mD$ and $0.2$ respectively with negligible capillary effects. Dry CO2 is injected at a constant reservoir volume rate of $50\,RB/D$ ($0.01\,PVI/D$ equivalent) through an injection well (shown in red line in **Fig. 6(a)**) completed only in the last three layers of the model. Brine saturation and initial pressure were kept constant at $1.0$ and $2500\,psia$ respectively at initial conditions. To imitate open aquifers and avoid excessive pressure buildup, pressure was maintained constant at the other end of the injector throughout the injection period. Supercritical CO2 is injected for a period of 10 days before injection well shut in, after which plume migration was monitored over a period of 1000 years.

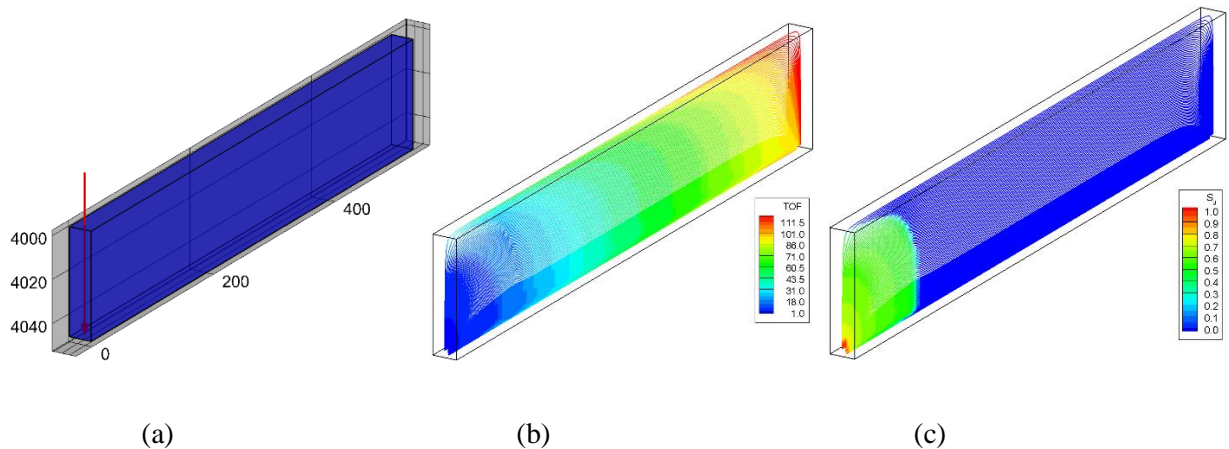

Figure 6: (a) Homogeneous 2D cross-section model showing the CO2 injection well in red line (b) Streamline distribution with Time of Flight contoured along each streamline (c) Saturation of CO2 rich phase at injector shut-in contoured along streamlines

**Figs. 6(b)** and **(c)** show the streamline distribution contoured with time of flight and CO2 rich phase saturation in at the end of CO2 injection period respectively. Solutions are mapped on the underlying grid and compared with the results from CO2STORE. CO2 rich phase saturations are compared in **Fig. 7** while aqueous phase CO2 mass fractions are compared in **Fig. 8**. Overall, our model showed good agreement with the commercial simulator. High values of CO2 rich phase saturations (and consequently, less than connate aqueous phase saturations) as well as low aqueous phase CO2 mass fractions in the vicinity of the injection completions indicate the impact of formation dry out which is duly captured by our streamline based simulation method. A well-studied feature of CO2 injection in saline aquifers is flow instability which results from an increase in aqueous phase density due to dissolution of CO2. This creates a vertical convection current which in turn causes fingering of CO2 bearing aqueous phase through virgin brine phase as shown in **Fig. 8**. Besides fluid phase and component concentration profiles on simulation grids, we compared CO2 injector bottomhole pressures (BHP) as a function of time shown in **Fig. 9**. A good agreement was also recorded between our approach and CO2STORE.

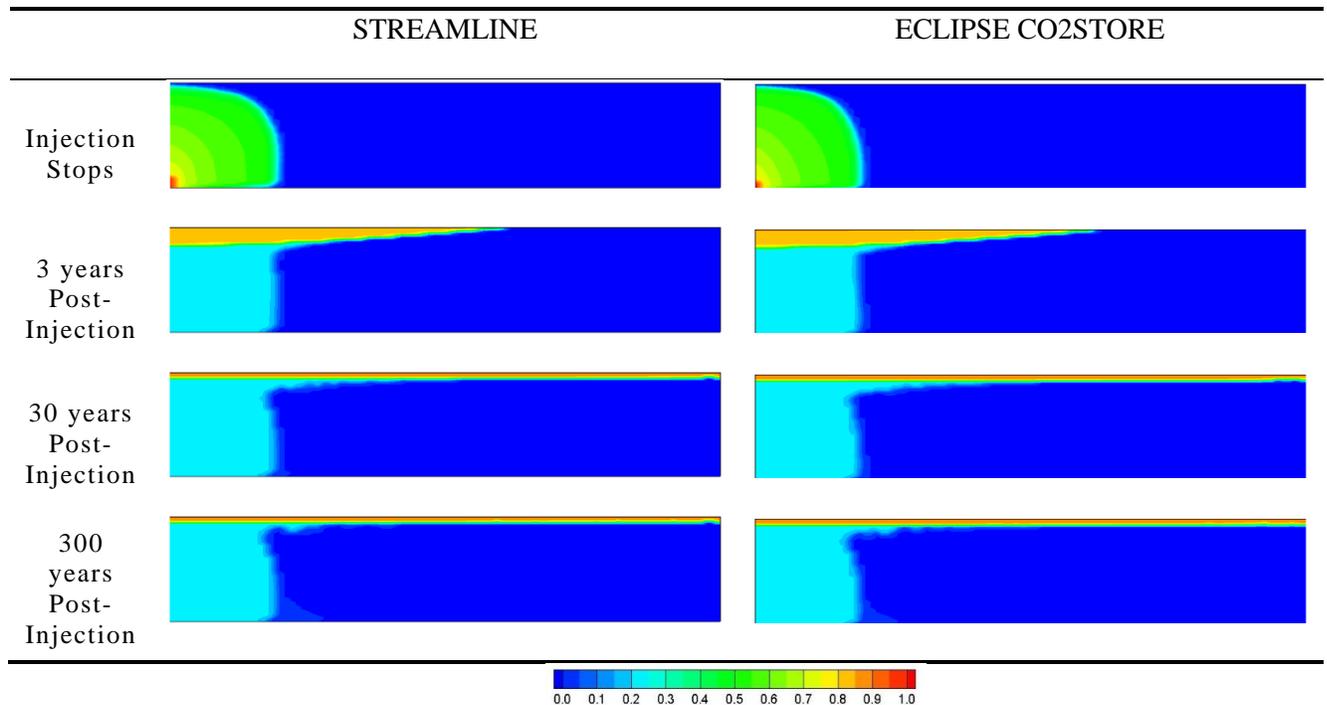

Figure 7: Time Lapse of CO2-rich phase saturation for 2D cross-section homogeneous model

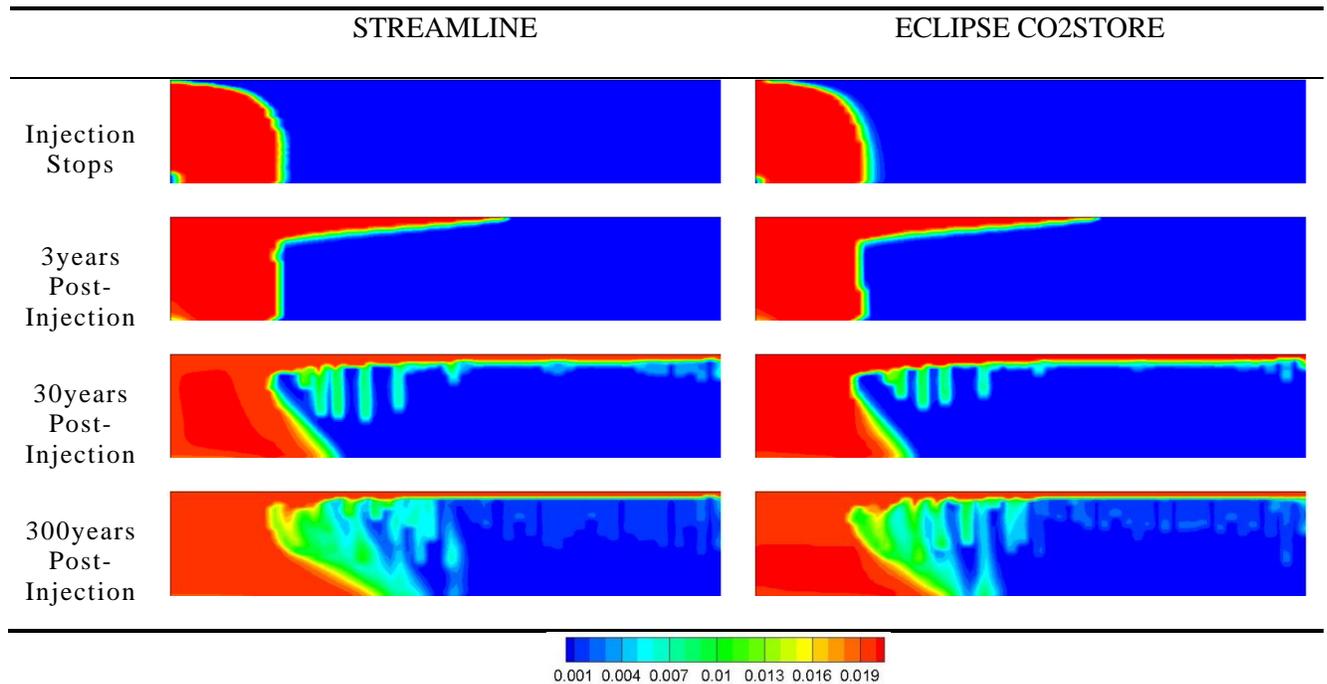

Figure 8: Time Lapse of CO2 mass fraction in the aqueous phase for 2D cross-section homogeneous model

Permeability and porosity variations with time through the porous media as a result of salt precipitation and mineralization are shown in **Fig. 10**. Note that rock porosity and permeability are altered mainly due to salt precipitation during injection period as effects of reaction that leads to CO2 mineralization is negligible at such time scale due to low reaction rates. Time scales of active CO2 trapping mechanisms can be clearly described through the life cycle plot shown in **Fig. 11**. After injector shut-in free CO2 plume continues to migrate and get trapped by residual mechanism. This leads to steady decrease in free CO2 saturation and increase in residual CO2 until the plume hits the other end of the formation in the constrained domain and can no longer migrate. Residual CO2 begins to decline due to dissolution in brine facilitated by gravity fingering and convective mixing. At 300 years, reaction and CO2 mineralization begin to take effect resulting in significant permeability and porosity reduction throughout the porous media as shown in **Fig. 10**.

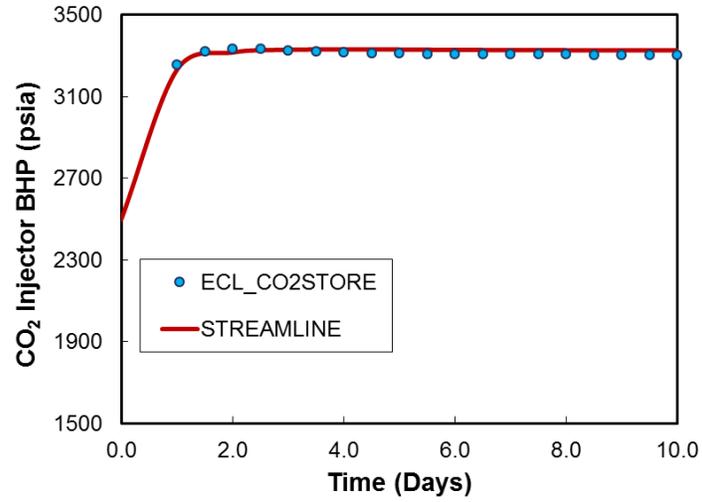

Figure 9: CO2 injector bottomhole pressure for homogeneous 2D cross-section homogeneous case

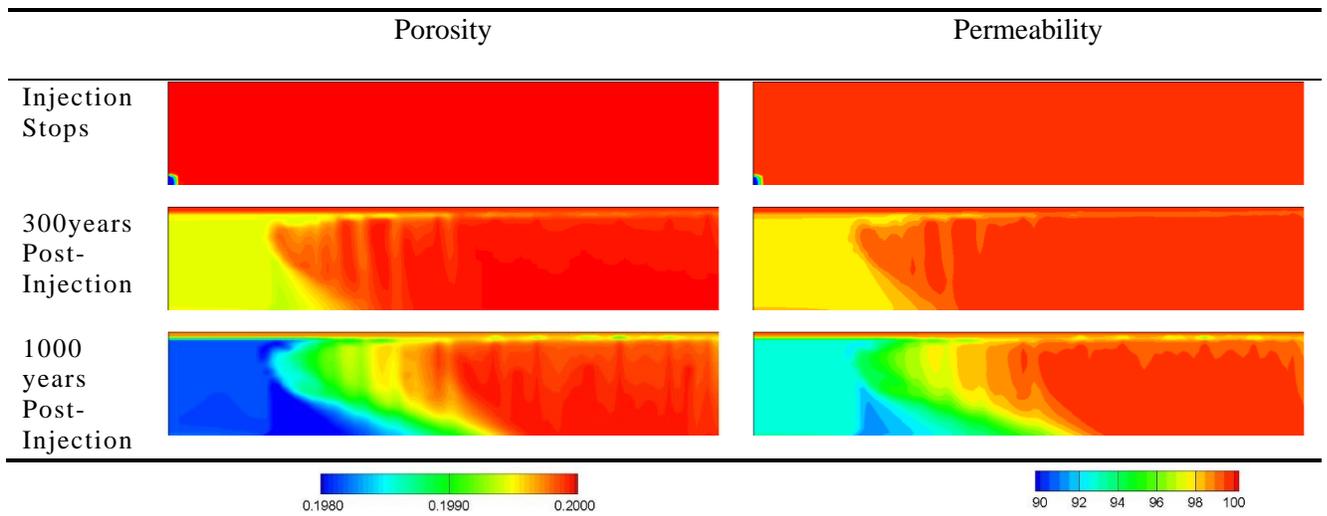

Figure 10: Time Lapse of grid porosity and permeability changes due to salt precipitation and reaction

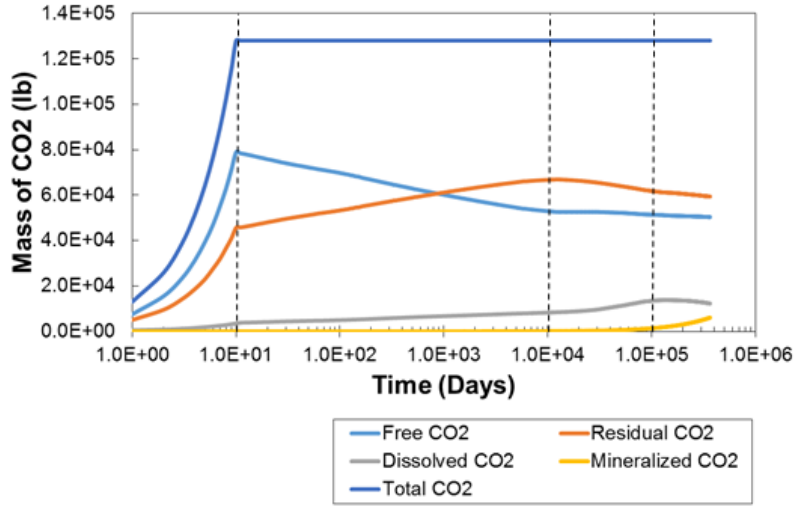

Figure 11: CO2 Life cycle over 300 years post-injection for homogeneous 2D cross-section homogeneous case

## 4.3 Example 3: Heterogeneous 2D Cross Section

A study was also conducted using a vertical cross section model with heterogeneous permeability field obtained from the first 35 (non-channelized) layers of the SPE 10 comparative model (Christie and Blunt, 2001). The flow domain is uniformly discretized into $150 \times 1 \times 35$ grid cells, each grid have similar dimensions as the homogeneous 2D cross section example discussed above. Grid permeability distribution shown (in natural logarithm) in **Fig. 12(a)** ranges between $0.02 - 8000 mD$ to add more geologic realism into the model. For simplicity, the formation is assumed incompressible while porosity takes a constant value of $0.2$ and capillary fluxes are negligible. Initial conditions are similar with the homogeneous 2D cross section example with initial pressures and brine saturations taking uniform values of $2500 psia$ and $1.0$ respectively. This model likewise consists of a CO2 injector at one end of the rectangular model and completed within the last 5 layers of the simulation grid. Similar boundary conditions of $50 RB/D$ ($\approx 0.01 PV/D$ equivalent) constant injection rate of CO2 and constant was imposed over $15 Days$ of CO2 injections. Again, the CO2 plume effects on formation was monitored over a period of 1000 years post-injection.

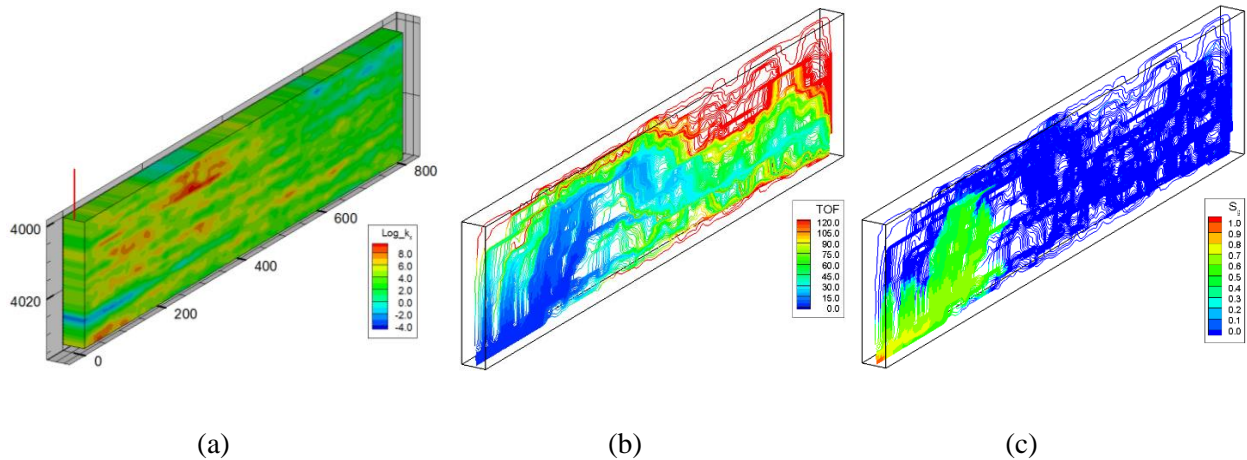

(a)            (b)            (c)

Figure 12: (a) Heterogeneous 2D cross-section model showing logarithm of permeability distribution and CO2 injection well in red line (b) Streamline distribution with Time of Flight contoured along each streamline (c) Saturation of CO2 rich phase at injector shut-in contoured along streamlines

An appealing feature of the streamlines methodology is an intuitive visualization of flow though connected volumes in a heterogeneous domain. The streamline distribution for this model with time of flight contours shown in **Fig. 12(b)** provides useful information not only about flow distribution, but also relative connectivity within the domain. Fluids typically flow preferential paths with lower times of flight along the streamlines. This can be easily observed by visual inspection of the CO2 rich phase saturation at the end of injection as calculated along the streamlines shown in **Fig. 12(c)**. Solutions on the streamline time of flight domain are mapped to the Cartesian domain for easy visual comparison with CO2STORE.

Temporal and spatial profiles of CO2 rich phase saturations and aqueous phase CO2 mass fractions on the finite difference grids are compared separately in **Figs. 13** and **14** respectively. Good agreement between the two simulations methods was recorded as well. Also the near-injection grid precipitation phenomena is well captured as indicated by unit saturations of the CO2 rich phase and lower mass fraction of CO2 dissolved in the aqueous phase. Injector BHP comparison, shown in **Fig. 15** likewise shows good agreement between our streamline based method and CO2STORE.

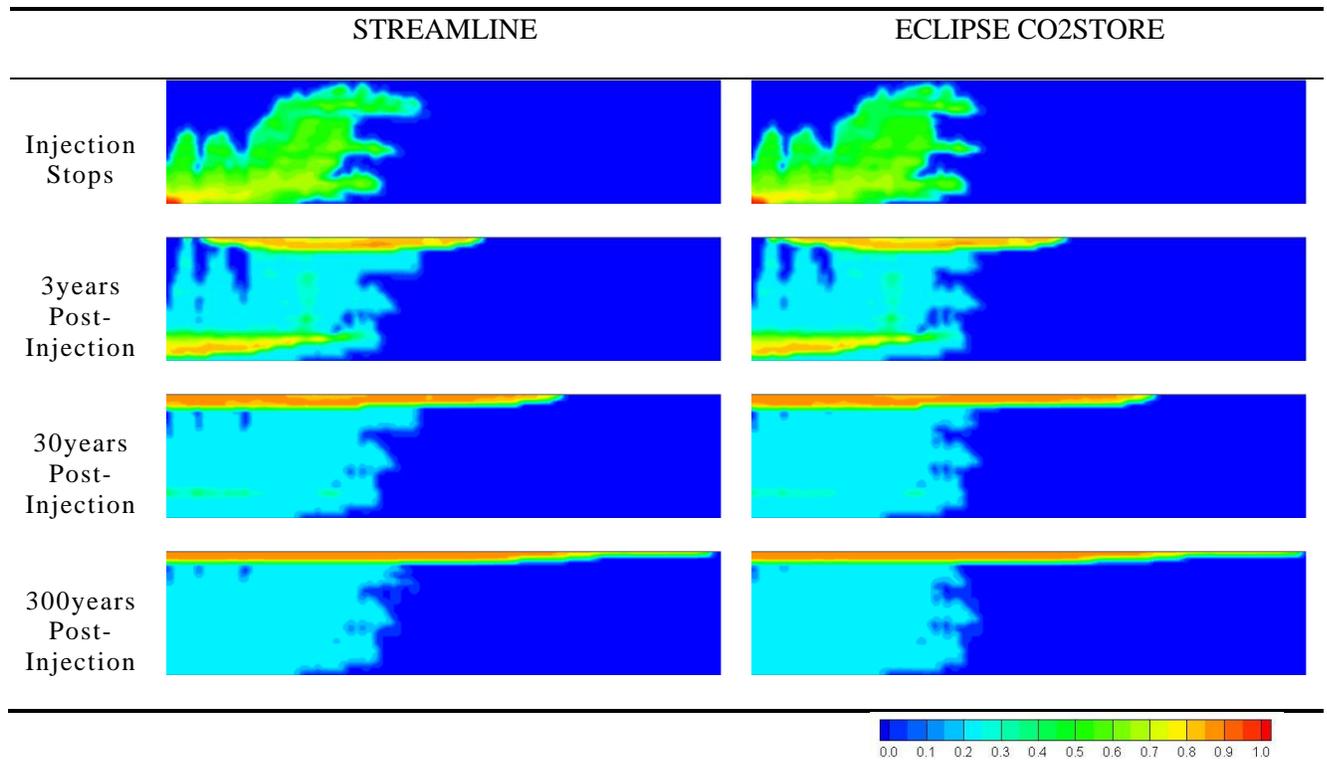

Figure 13: Time Lapse of CO2-rich phase saturation for 2D cross-section heterogeneous model

For this example, a CO2 life cycle plot is provided in **Fig. 16** as a graphic representation of the variations and time scales of CO2 trapping mechanisms over $1000\, years$ post-injection. Comparing this plot with **Fig. 11**, it can be observed that the point of maximum residual trapping of the CO2 is delayed in the heterogeneous model compared with the homogeneous model. One explanation for this is an improvement in vertical sweep due to flow barriers which delay gravity override in the heterogeneous model even though comparable volumetric rates of CO2 were injected in both cases. Consequently, higher residual trapping results with greater reduction in free CO2 structurally trapped at $1000\, years$. Time scales for solubility and mineral trapping are more or less comparable in both the homogeneous and heterogeneous examples, except for slight delay in the onset of convective mixing in the heterogeneous case.

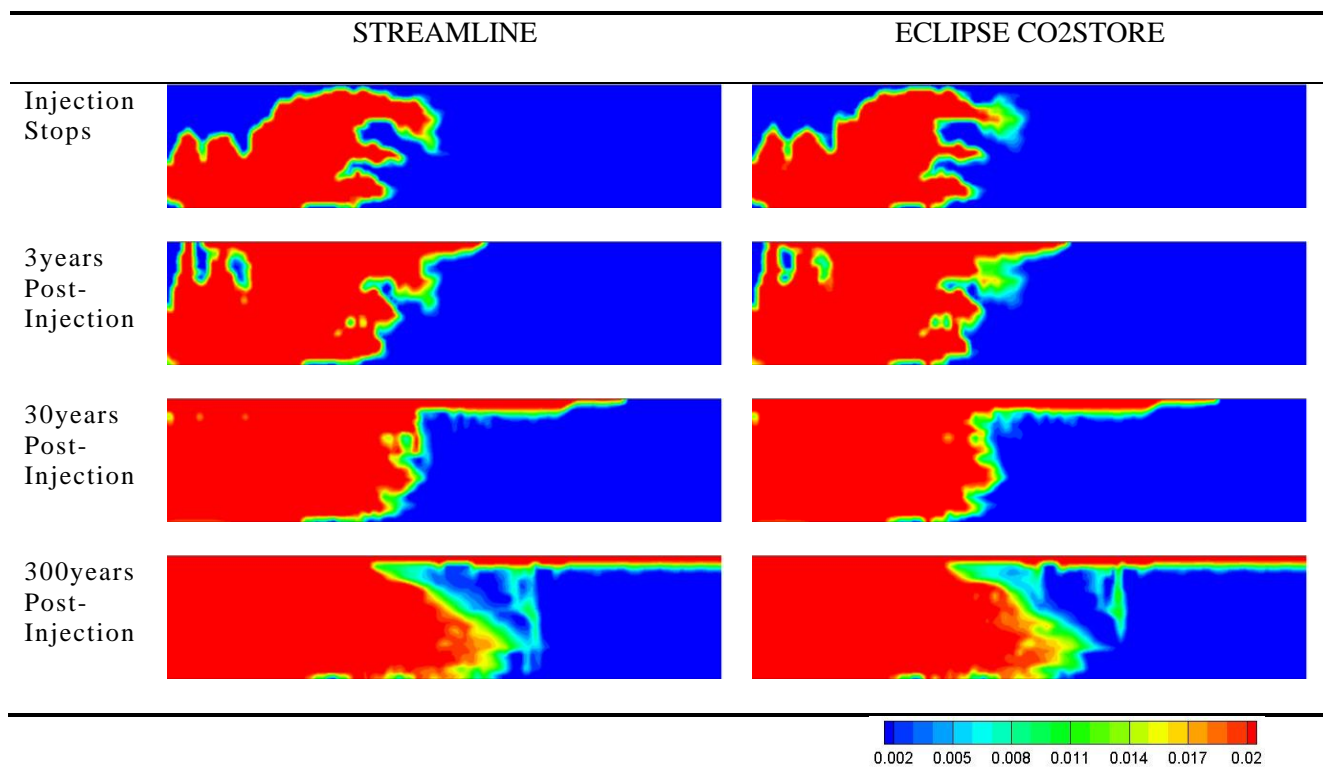

Figure 14: Time Lapse of CO2 mass fraction in the aqueous phase for 2D cross-section heterogeneous model

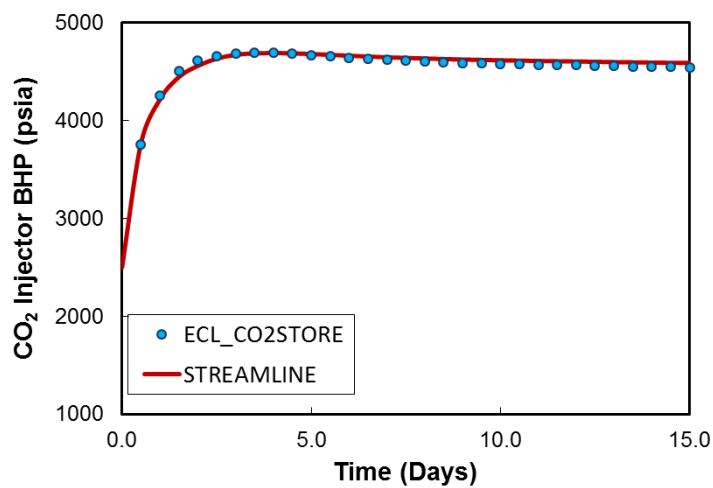

Figure 15: CO2 injector bottomhole pressure for heterogeneous 2D cross-section case

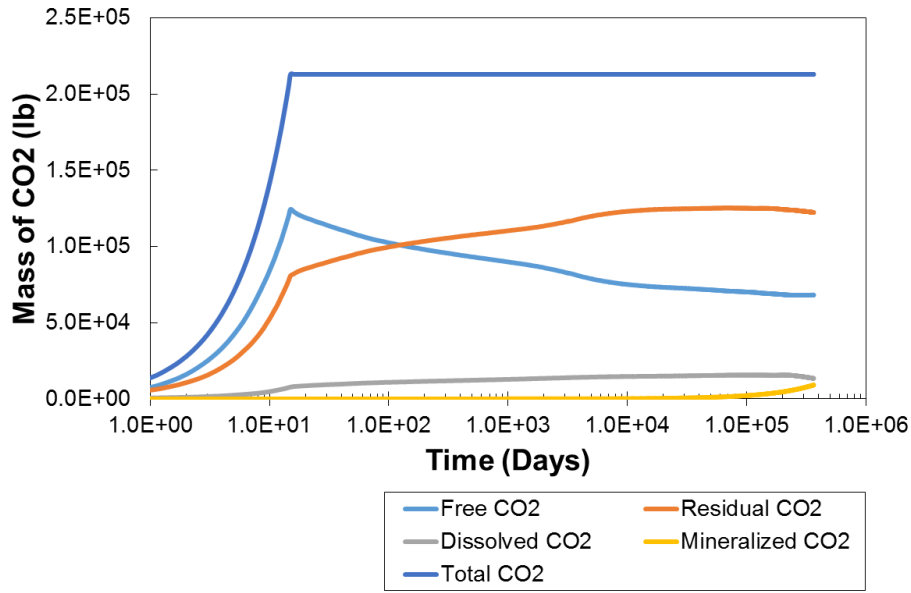

Figure 16: CO2 Life cycle over 300 years post-injection for heterogeneous 2D cross-section case

# 5  Discussion of Results

In this section we present a technical summary of the results so far including the computational benefit of our streamline-based methodology compared to conventional finite volume methods. We also provide a description of the improvement offered by our method over existing streamline based methods for CO2 sequestration applications.

## 5.1  Value of Streamline Simulation

The subsurface environment is typically heterogeneous and quick assessment of storage potentials of saline aquifers requires efficient analytic and flow simulation approaches. Streamline simulation has been extensively studied and shown to effectively handle large and small scale heterogeneities while avoiding grid orientation and numerical artefacts. Consequently, streamline simulation potentially offers benefits in improved tracking of CO2 plume during injection. In all examples presented so far, we have consistently shown reasonable agreement between our simulation approach and commercial finite difference simulator based on standard flow dynamics metrics including spatial profiles of phase saturations, component mass fractions and injector BHP variations with time. Here we investigate the value of our streamline-based simulation methodology in terms of convergence and computational advantage.

### *5.1.1 Convergence Studies*

In this section, we compare simulation convergence performance between the regular streamline methodology, multiscale streamlines approach and finite difference simulation. For both homogeneous and heterogeneous 2D cross-section models conducted series of numerical experiments at various grid resolutions to investigate the rate of convergence of each simulation approach. For both cases, we start with a $40 \times 5$ grid model with $50 \times 50 \times 20 \, ft$ homogeneous grid dimensions and vertically refine by a factor of 2 twice to obtain additional models of vertical grid dimensions $10$ and $20$. The reference model was set to have a vertical grid dimension of $80$. An infinite aquifer domain was mimicked with pore volume multiplier value of $3000$ at both ends of the model. We imposed a constant initial reservoir pressure of $2500 \, psia$ for both models and CO2 is injected at $200 \, RB/D$ over a period of $200 \, days$. A homogenous porosity value of $0.2$ was applied with permeability value of $100 \, mD$ for the homogeneous model, while the natural logarithm of permeability for the heterogeneous model ranges from $-6$ to $6$.

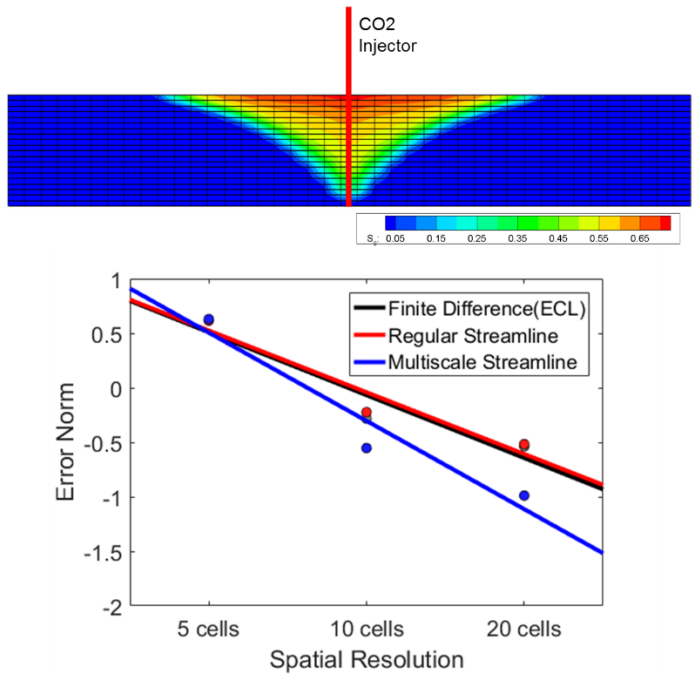

Figure 17: Grid resolution study for homogeneous 2D cross-section case showing the gas saturation profile for the 20-cell vertical resolution case (top) and the grid convergence comparison plots (bottom)

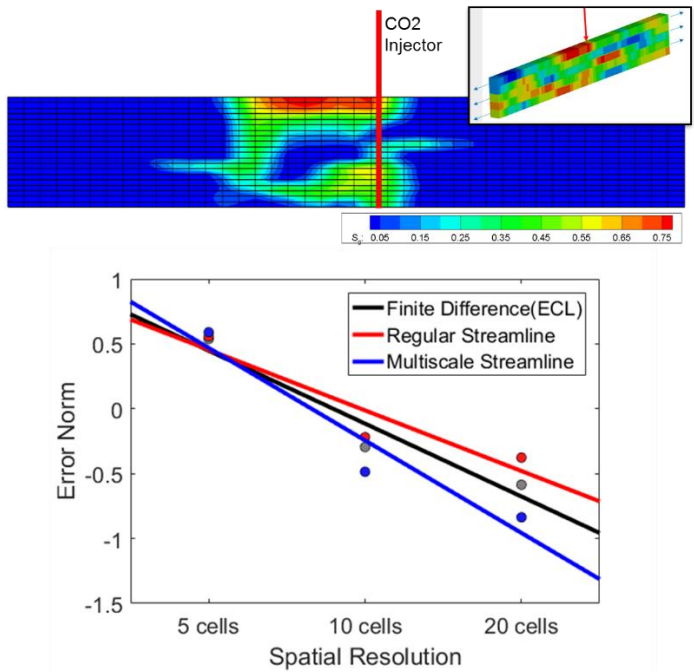

Figure 18: Grid resolution study for heterogeneous 2D cross-section case showing the gas saturation profile for the 20-cell vertical resolution case (top) and the grid convergence comparison plots (bottom). Inset shows the ln(permeability) distribution on the coarse 5-layer grid

The results and observation obtained from the numerical experiments is summarized in **Fig. 17** and **18**Error! Reference source not found. for the homogeneous and heterogeneous models respectively. For all three simulation methods, we summarize the grid resolution experiment with a plot of the error norm, computed as $\log_{10}\left(\left\|s_g - s_{g_{REF}}\right\|\right)$, is plotted against the number of vertical grid cells. The error is simply the L2 norm of the difference between the calculated gas saturation on the low resolution grid $s_g$ and the reference gas saturation computed on the 80-layer grid $s_{g_{REF}}$. An improved convergence was observed with the multiscale streamline simulation methodology for both homogeneous and heterogeneous models. This is seen to be a direct implication of the higher resolution $\tau$ discretization which also allows for gravity corrections on the refined grid, thereby minimizing the smearing effect that results from grid-to-streamline mappings during streamline simulation.

### *5.1.2 Time Stepping*

As shown in **Fig. 19**, in all the three cases number of required time steps is reduced by a factor of 10 on the average compared to CO2STORE. This may be attributed to the unique effectiveness of streamline based methods at handling convective transport leaving a fair amount transverse flux corrections per predictor step. Number of iterative IMPES iterations, is thus small for reasonably high injection rates as those practical for CO2 sequestration applications. We however recognize, as a limitation, that this benefit of streamline based approach may diminish as transverse fluxes increase with gravity and/or capillary dominated systems.

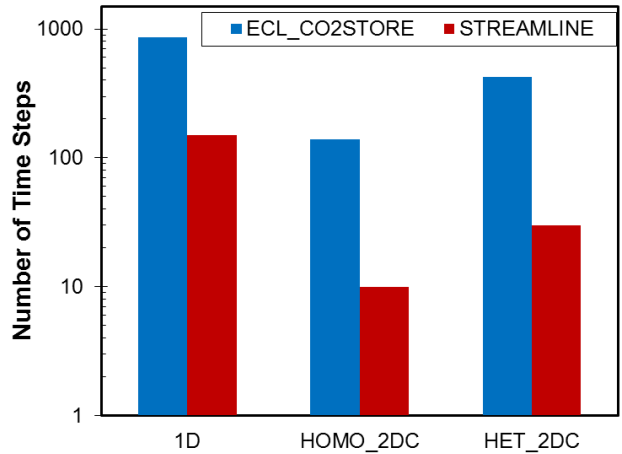

Figure 19: Comparison of required number of time steps

## 5.2 Fluid Compressibility Effects

The strengths of streamline based simulation and flow description in heterogeneous porous media have previously been exploited for CO2 sequestration modeling by few authors (Obi and Blunt, 2006; Qi et al., 2009). These streamline models were however built on the assumption of incompressible fluids, which may not be quite accurate for the CO2 rich phase. Aside the fact that ignoring compressibility effects may lead to inaccurate pressure buildup estimation in the aquifer, studies have clearly shown that compressibility effects have significant impact on the eventual size and shape of the CO2 plume during injection (Vilarrasa et al., 2010). A possible explanation for this is the non-linear relationship that exists between pressure, CO2 rich phase and the aqueous phase mass fraction of CO2. High pressures results in larger equilibrium concentration of CO2 in the aqueous phase, which in turn affect the system of mass balance equations which solves to give the pressure distribution. Therefore, since the 1D streamline grids are generated based on the total fluxes computed from the pressure distribution, disregarding compressibility effects becomes detrimental to realistic representation of the flow physics.

We illustrate this with two numerical experiments were conducted with separate boundary conditions using the 1D model. We have chosen the simple model so that buoyancy effects do not interfere with compressibility effects, however as we will show later using a field example, similar conclusions can be reached with 3D geologically realistic cases. For both cases, the initial formation pressure is $3000\,psi$ and, again a constant pressure boundary was maintained at the opposite end of the injection well to mimic and open reservoir boundary. The injection well was constrained at a constant pressure of $5500\,psi$ for the first experiment, while CO2 was injected as a constant reservoir rate of $1.7\,RB/D$.

The compressibility effects are presented for the BHP constrained and rate constrained cases in **Figs. 20 a**nd **21** respectively. For the constant injection BHP case, it is expected that the CO2 rich phase saturation shock front for the compressible case will be ahead of the case where compressibility is not considered. This is because compressibility acts here as an additional source where the flux divergence is positive. It also shows up as a higher grid pressure at locations where CO2 rich phase saturation is non zero as shown in **Fig. 20**. Higher surface rate injection therefore results for the compressible case, which imply higher infectivity and ultimately, larger storage capacity than the incompressible case. For the constant reservoir injection rate constraint experiment, the CO2 plume apparently migrates faster in the incompressible case at $1.5\,days$ as shown in **Fig. 21**. This seems contradictory, but it turns out that the lag in the compressible case is due to the lower injection BHP at early time, during which compressibility acts as a local sink and pressure slowly builds up in the reservoir. Higher surface injection rate, and thus, high storage capacity is however achieved with the compressible case again in this experiment due to compressibility effects. Concretely, both experiments show that ignoring compressibility effects, as in previous streamline-based models, under-estimates the storage capacity by a factor of 15%.

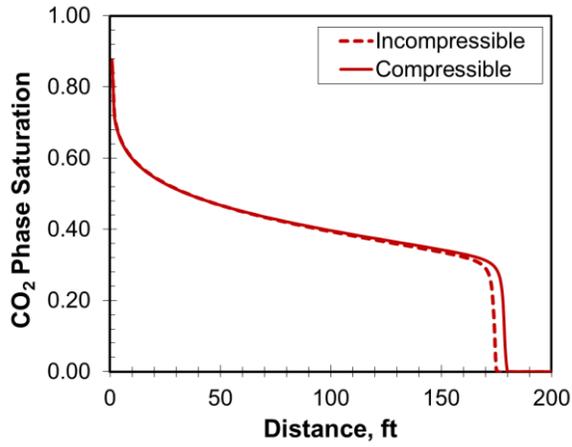
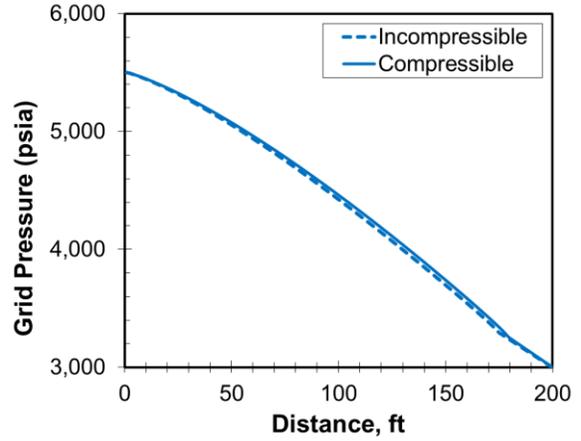

(a)                          (b)

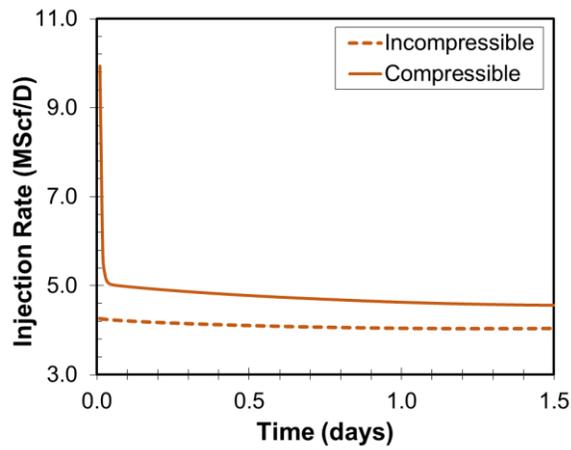

(c)

Figure 20: Simulation responses for a constant injection pressure case showing (a) CO2 phase saturation profile, (b) cell pressure profile and (c) CO2 surface injection rate for the constant bottomhole pressure case

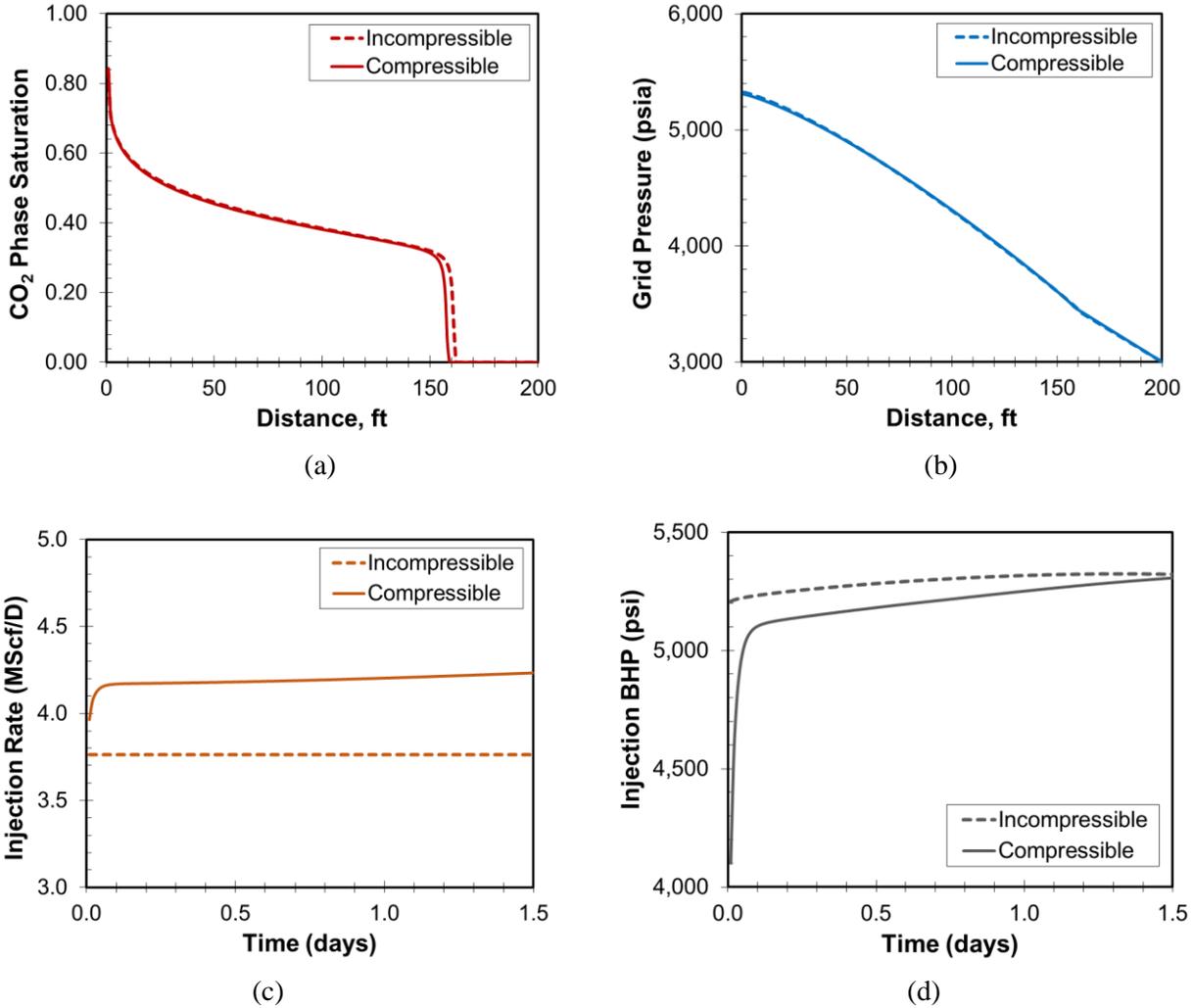

Figure 21: Simulation responses showing (a) CO2 phase saturation profile, (b) cell pressure profile, (c) CO2 surface injection rate and (d) Injector bottomhole pressure response for the constant subsurface volume injection rate case

## 5.3 Formation Dry-Out Effects

On another note, formation dry out effects due to salt precipitation in the near injection well region, as shown by previous studies (Giorgis et al., 2007; Pruess and Müller, 2009), has significant negative influence on well infectivity over time. This poses a significant challenge to the overall operation of CO2 sequestration projects and as such, incorporating this effect in simulation models becomes important. Solubility calculations based on modified Redlich-Kwong equation of state (Spycher and Pruess, 2005) was included in the recent streamline model (Qi et al., 2009) but with more emphasis on the aqueous phase components, which turned out to ignore formation dry out effects. Besides that, solubility calculations were carried out after mapping convective transport solutions from the streamline domain onto the finite difference grids. This approach suffers from a smearing effect whereby component distribution between phases is more or less averaged over a set a grid cells as decided by the pressure update time step. This makes it difficult to fully capture formation dry out effects except with very frequent pressure updates. In our streamline based approach, mutual solubility calculations are carried out in tandem with convective transport before mapping to the grid. This ensures local precipitation effects are better captured without resorting to small time steps.

We demonstrate this effect by comparing the grid-based and streamline-based mutual solubility calculations with the homogeneous 2D cross section but under two different conditions: with and without capillary fluxes. For the case with capillary pressure we applied the van Genuchten type capillary pressure function (Van Genuchten, 1980) expressed as:

$$p_c = p_g - p_{aq} = p_0 \left( \tilde{S}^{\left(\frac{n}{1-n}\right)} - 1 \right)^{\frac{1}{n}} \; ; \quad \tilde{S} = \frac{s_{aq} - s_{aq,c}}{1 - s_{aq,c}} \tag{48}$$

Where $s_{aq,c}$ denote aqueous phase connate saturation, $p_0 = 0.1$ and $n = 1.31$. All other conditions such as discretization, well locations and completions and injection rates are kept the same as described in the homogeneous 2D cross section example. Average solid saturation and permeability reductions in the injection grids are tracked for both cases over the simulation period of $10 \, days$. The results of this experiment are presented in **Figs. 22(a)** and **(b)** for cases without and with capillary effects respectively. The general trend is an increase in solid saturation and reduction in grid permeability due to salt precipitation. For the case without capillary effects, precipitation in the injection grid stops at 4days making solid saturation and grid permeability stay constant for the rest of the simulation period. At this point injection grids have effectively dried out with unit gas phase saturations. On the other hand under the influence of capillarity as in the second case, the injection grids do not dry out within the simulation time frame. This is because the strong water-wet nature of the rock causes more brine to be drawn to the injection grids as water continually gets vaporized into the CO2 rich phase. As a result, solid saturation increases while grid permeability reduces dramatically compared to the case without capillary effects. This is a reproduction of the effects previously studied by other authors (Giorgis et al., 2007; Kleinitz et al., 2003; Ott et al., 2015; Peysson et al., 2014) but with streamline based methods.

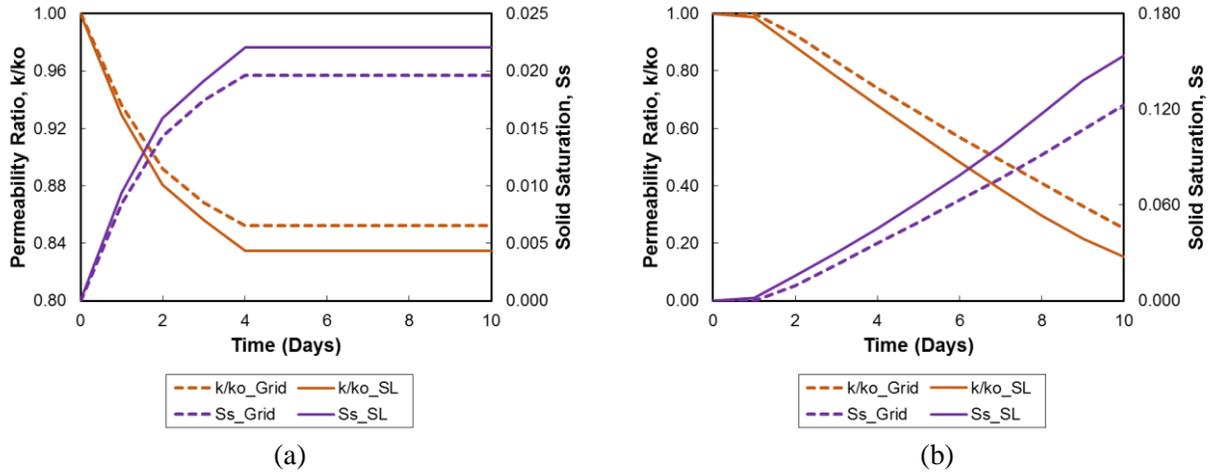

Figure 22: Comparison of permeability and porosity changes obtained from grid-based and streamline-based mutual solubility calculation models (a) without capillarity and (b) with capillarity

Equally important is the observation from the comparison between grid-based and streamline-based mutual solubility calculations for both cases with and without capillary effects. The previously discussed smearing effect associated with the grid-based approach results in poor capture of the near well region precipitation phenomena. This is indicated as smaller increase in solid saturation and smaller reduction in grid permeability when the grid dries out. Without capillary forces, there are mild differences between the two solubility calculation approaches with about 10% and 2% differences in solid saturation and permeability reduction respectively. As formation get more and more water wet, the weakness of the grid-based approach in capturing the near well details get more significant with about 25% and 40% differences in solid saturation and permeability reduction respectively. Clearly such disparity in near well grid permeability results in differences in well injectivities which in turn can impact the evaluation of the aquifer storage potential and/or overall project viability.

# 6 Field Example: Johansen Case

## 6.1 Background on Field Model

Finally we applied our streamline-based CO2 sequestration modeling approach to the Johansen field (Eigestad et al., 2009), one of the world's few aquifers considered for large scale storage of anthropogenic CO2. The Johansen formation, located offshore the Norwegian coast, was intended to serve two gas fired power plants in its vicinity having combined CO2 emissions of over 3Mt/year. The Johansen formation exists at depths between $7,200$ and $10,000\,ft$ below sea level, therefore with large overburden pressure which makes it suitable for CO2 Storage. The geological model indicates layers of sand and shale deposition with average thickness of over $300\,ft$.

A sector model of the field was publically made available with $100 \times 100 \times 11$ grid discretization with average cell dimensions of $1,500 \times 1,500 \times 70\,ft$. The Johansen sand is practically 'sandwiched' by the Dunlin shale in the top 5 grid layers and the Amundsen shale in the last grid layer. For the purpose of our simulations, we assume these shale zones are sealing and are therefore the respective simulation grids are considered inactive. **Fig. 23** presents the model structure and grid properties (ignoring shale layers) with permeability ranging between $64 - 1,660\,mD$ and porosity between $10 - 28\%$. The major fault existing in the middle of the formation is assumed to be sealing due to its large throw. All other faults in the formation are assumed to be sealing, and thus assigned zero transmissibility multipliers.

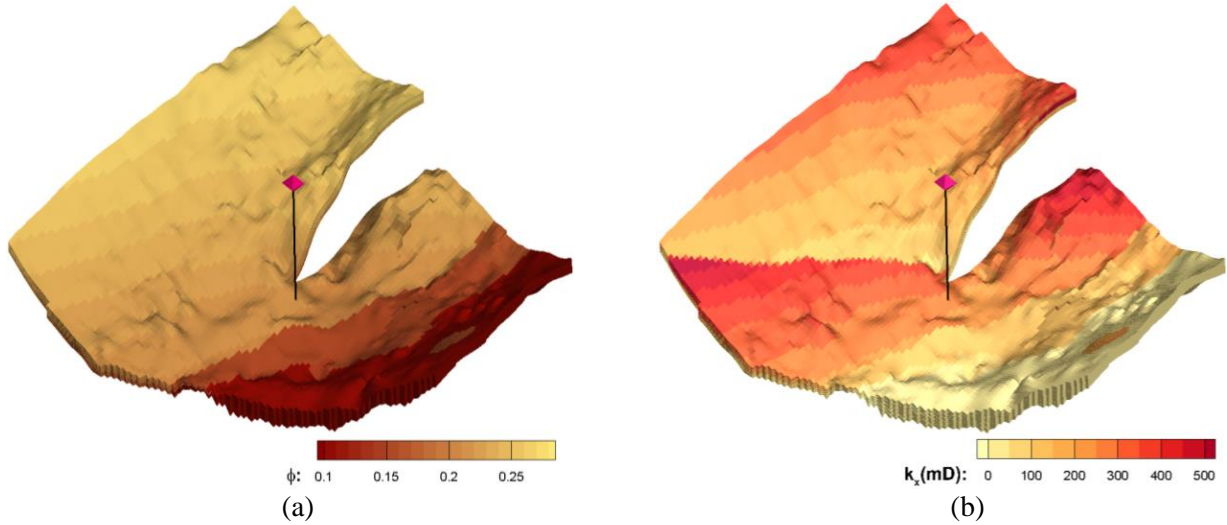

Figure 23: (a) Grid porosity and (b) Logarithm of grid permeability distribution of Johansen model

## 6.2 Numerical Simulation Set-up

The simulation model consists of a single well located in the center of the formation, similar to Eigestad et al. (2009). The well, assumed to connect to all 5 sand layers with a wellbore radius of $0.5\,ft$, injects supercritical CO2 at a rate of $483\,MMscf/D$ into the formation. This is equivalent to $29,000\,tons/day$ or $10\,Mtons/yr$. We applied a pore volume multiplier of $2,000$ in boundary cells to mimic a continuous aquifer. CO2 injection period lasted for $30,000$ days ($\approx 82$ years) and CO2 plume was monitored over a period of $2,000$ years post-injection.

A streamline based simulation model was conducted on the model with $1,000$ streamlines sufficient to span the domain. Pressure distribution shown in **Fig. 24(a)** indicate significant amount of pressure gradient (up to $2,000\,psia$) exists owing to large structural variations in the aquifer model. Streamline distribution at the end of the injection period is shown in **Fig. 24(b)** with the $\tau$ contoured along each streamline. Phase saturations and component mass fraction results from our streamline-based simulation approach was compared with the results from CO2STORE in **Fig. 25**. Again, good agreement was obtained between the two methods. Formation dry out effects are clearly noticeable, as indicated by unit CO2 rich phase saturations together with corresponding high aqueous phase salt concentration and low concentration of dissolved CO2 in the vicinity of the injection well. CO2 injector BHP also shows good agreement with the commercial simulator as shown in **Fig. 26**.

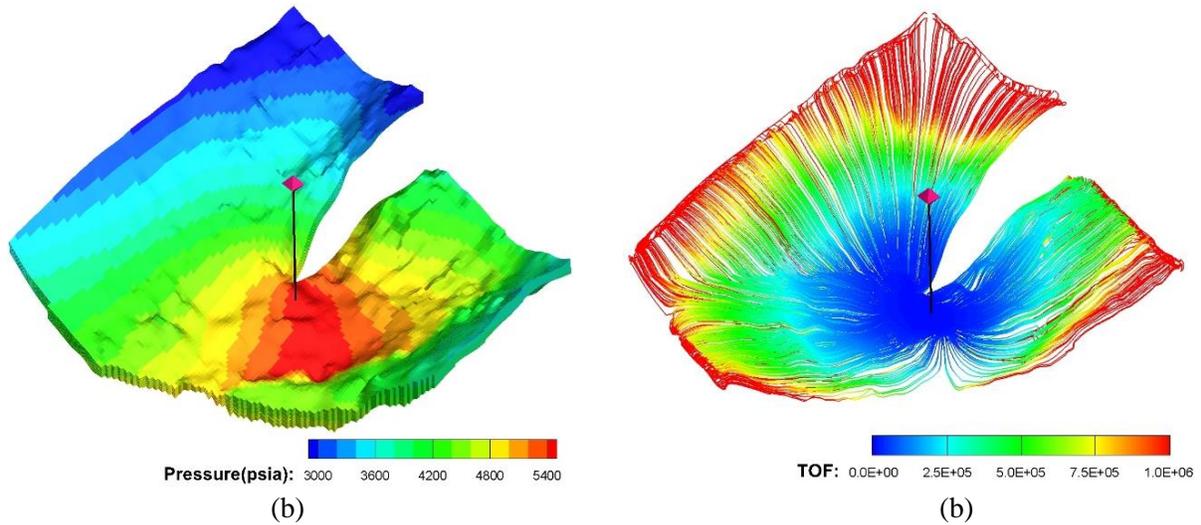

Figure 24: (a) Grid pressure and (b) Streamline distribution with time of flight contours at Injector shut-in

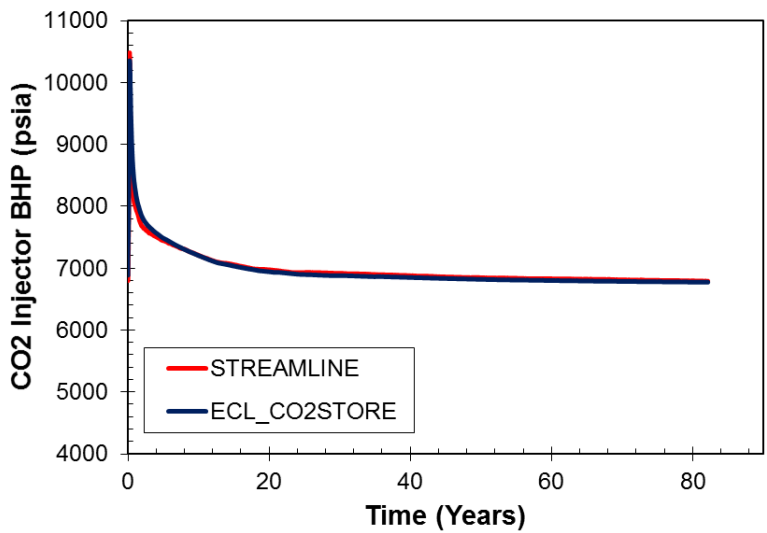

Figure 25: CO2 Injector BHP comparison between streamline-based simulation and CO2STORE for Johansen case

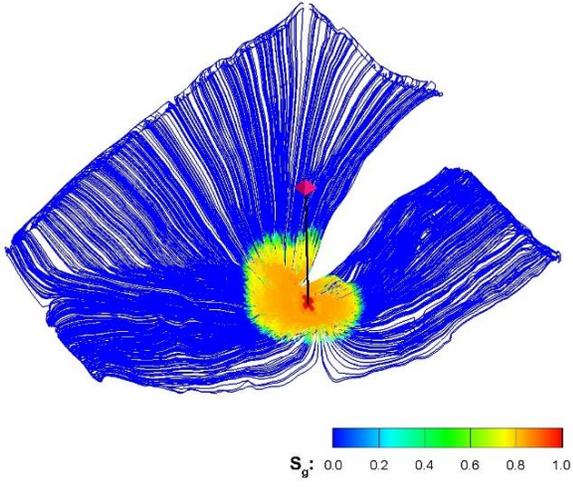
(a)

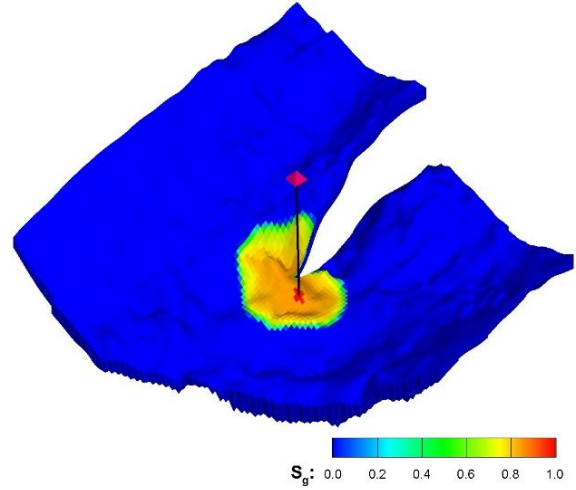
(b)

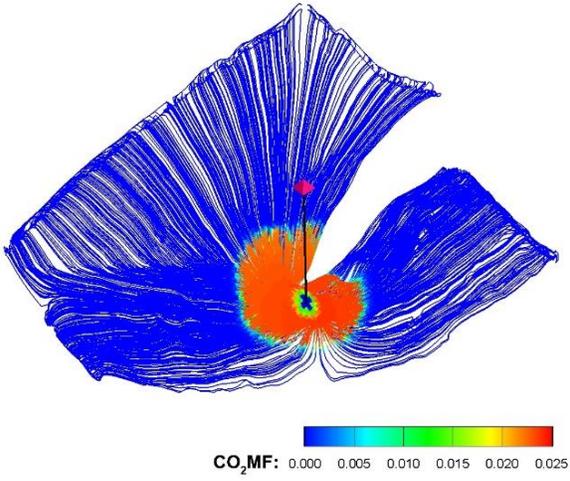
(c)

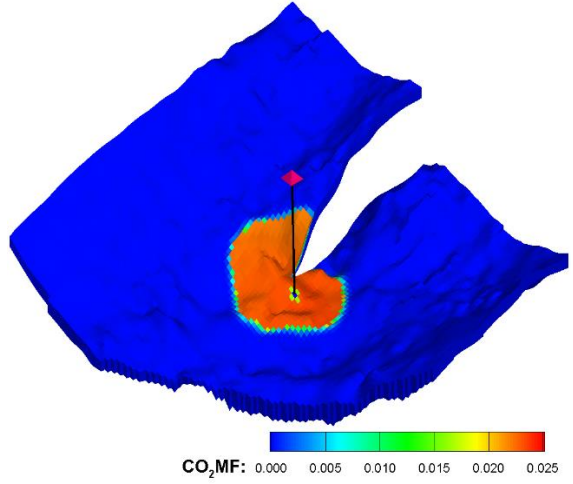
(d)

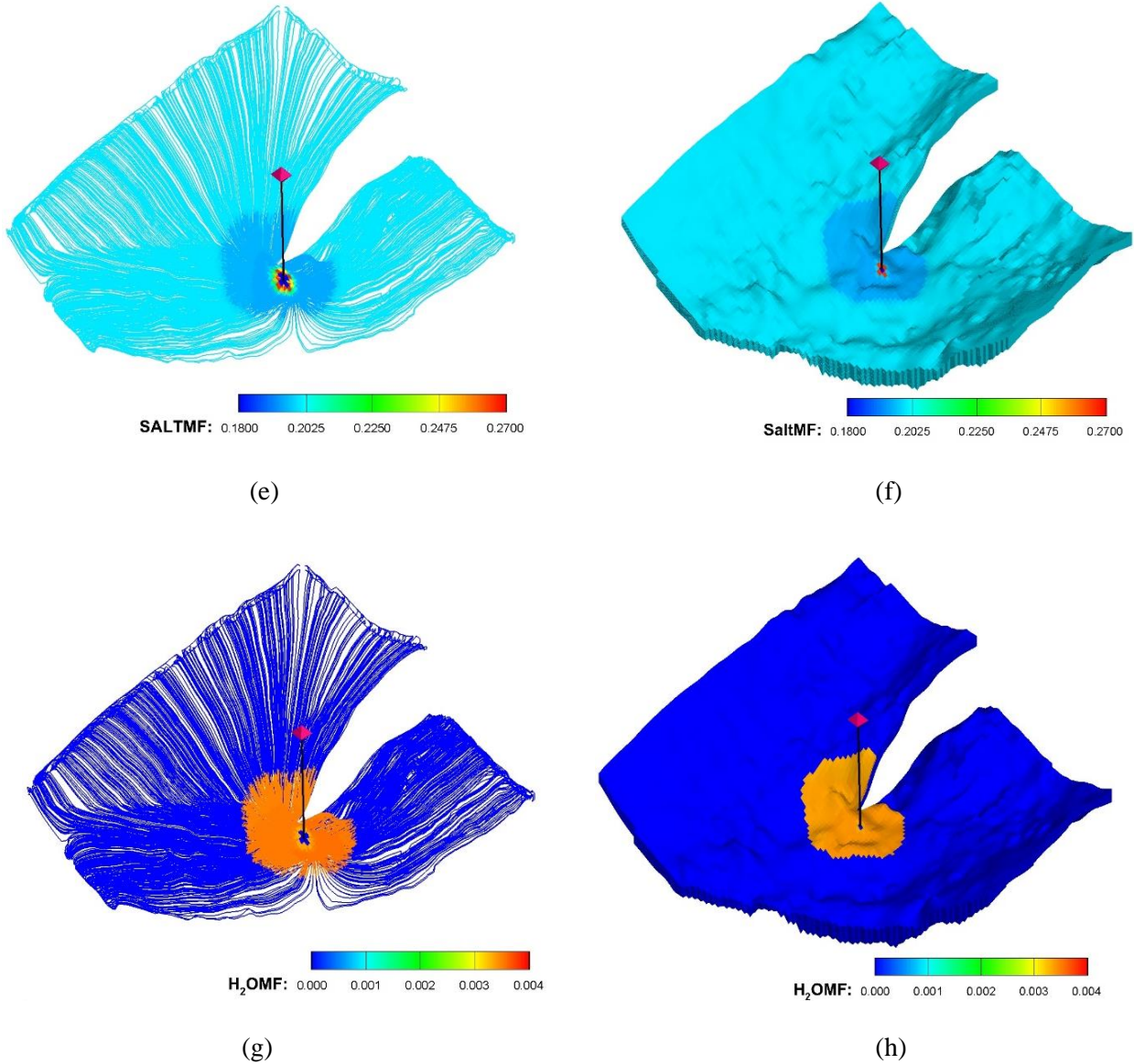

Figure 26: CO2 rich phase saturation from (a) streamline and (b) CO2STORE; aqueous phase CO2 mass fraction from (c) streamline and (d) CO2STORE; aqueous phase salt mass fraction from (e) streamline and (f) CO2STORE and finally, water mass fraction in the CO2 rich phase from (g) streamline and (h) CO2STORE at the end of gas injection

## 6.3 Fluid Compressibility Effects

Again, fluid compressibility effect was replicated for the Johansen case. Here CO2 injection was constrained at a constant bottomhole pressure value of $6000\, psia$ over an injection period of $80\, years$.

In **Fig. 27(a)**, we overlay plots of five saturation contours each for both compressible and incompressible cases on the finite difference grid. It turns out, consonant with the 1D pressure constrained injection case presented in Fig. , that the incompressible saturation front mostly lags the compressible one. Similar explanation suffices for this observation. During injection fluid compressibility effects acts as a pseudosink since flow divergence in the vicinity of the injection well becomes less than zero. As a result, larger fluid volumes at surface and reservoir conditions invade the reservoir, as confirmed with the surface injection rate plots in **Fig. 20**.

In addition however, for realistic fully 3D models where gravity override is important, fluid compressibility play a significant role in plume migration at the lead edge of the $CO_2$ plume directly beneath an impermeable seal. Therefore putting it all in context, ignoring fluid compressibility effects in streamline simulation of $CO_2$ injection in saline aquifers results in under-estimation of the extend of plume migration during $CO_2$ injection. Also, in agreement with the estimate made for the 1D simulation case, approximately $15\%$ reduction in aquifer storage capacity results from ignoring fluid compressibility effects.

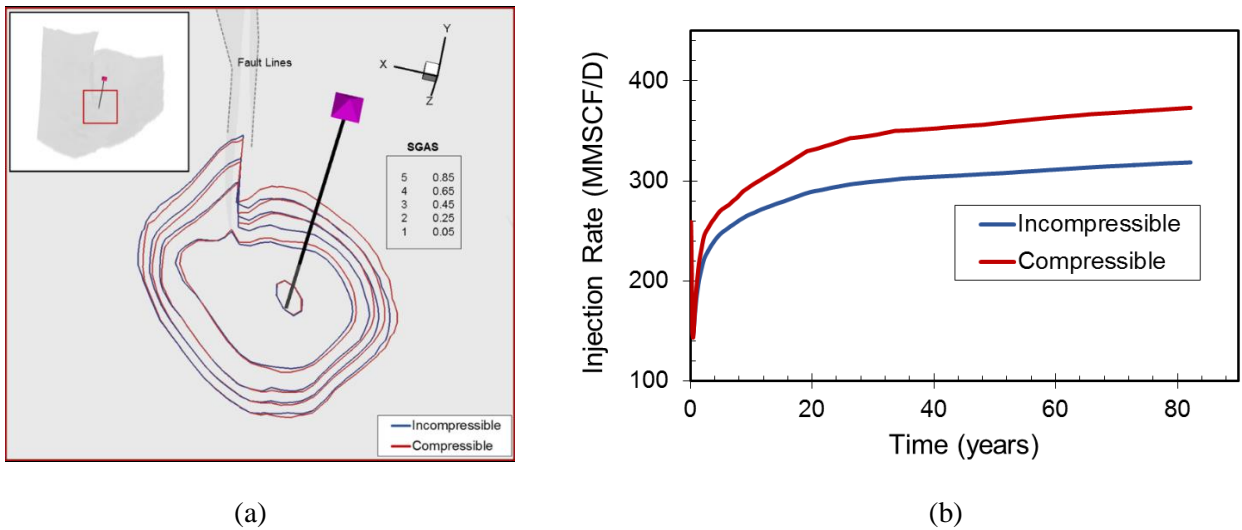

(a)          (b)

Figure 27: Plots showing fluid compressibility effects with respect to (a) $CO_2$ plume migration and (b) surface injection rate of $CO_2$

## 6.4 Life Cycle of CO2

The post injection period is characterized by upwards migration of CO2 plume under the influence of gravity forces, in the process of which more residual CO2 is trapped as shown in the life cycle plot in **Fig. 28**. A closer look at the CO2 rich phase saturation profile at 2,000 years post-injection in **Fig. 29(a)** reveals structural trapping whereby pools of CO2 rich phase are immobilized at saturations higher than residual saturation values due to the geometry of the surface of the formation immediately below the seals. Note that the amount of structurally trapped CO2 is included in the quantity of free CO2 plotted in **Fig. 28** and not specifically quantified. Assuming these caprocks are perfectly sealing, the structurally trapped pools of CO2 gradually reduce in size due to convective mixing that is generated over time.

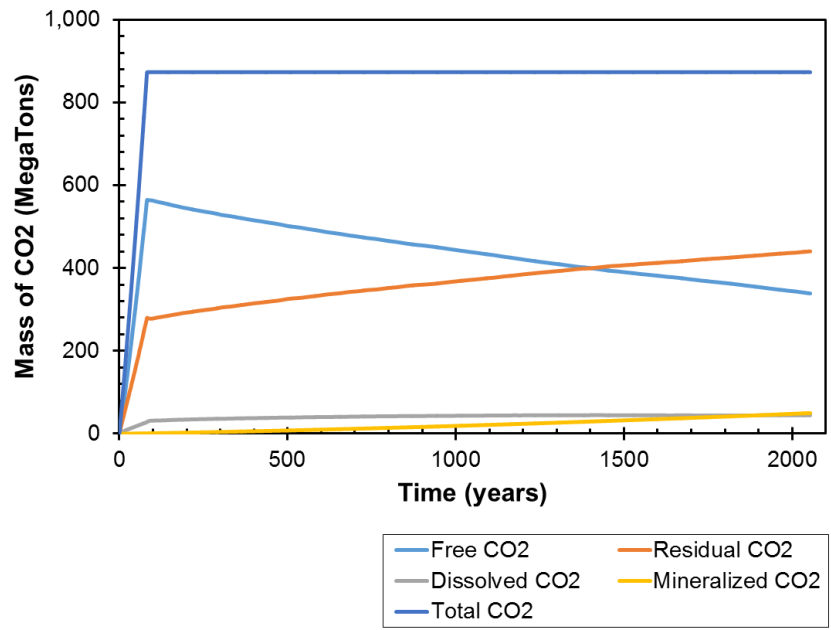

Figure 28: CO2 Life cycle for the Johansen case

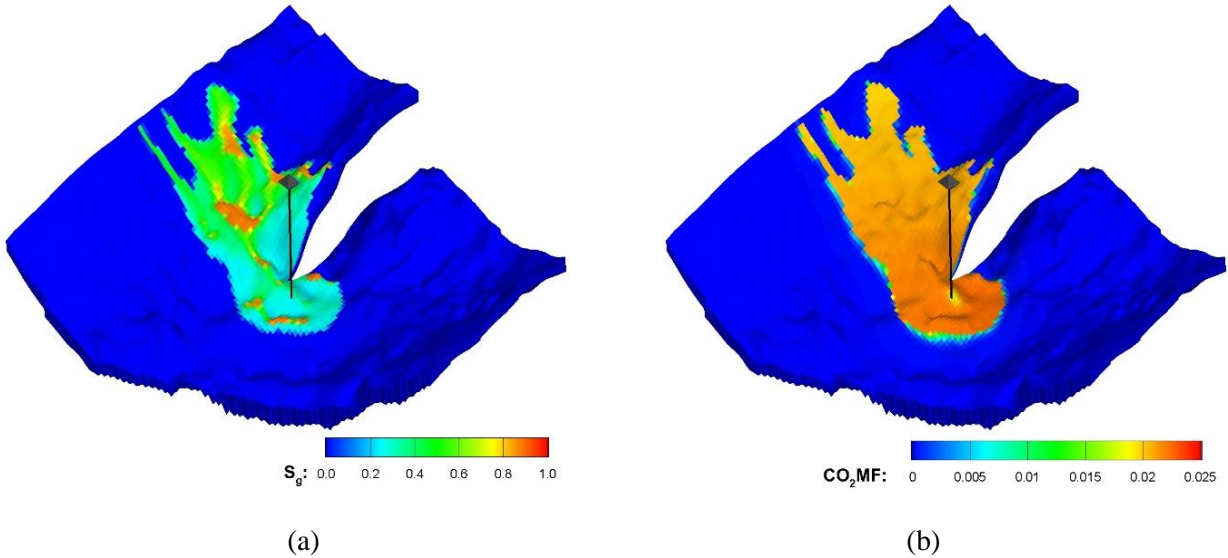

Figure 29: (a) CO2 rich phase saturation and (b) Aqueous phase CO2 mass fraction at 2000 years post-injection

# 7  Concluding Remarks

A streamline-based simulation approach for CO2 sequestration in saline aquifers has been presented. The more approach improves on previous streamline models to account for pertinent physics which are vital for correct and reliable prediction of the multiphase flow during CO2 storage process. Our method is based on an iterative IMPES scheme in which fluid properties, phase saturations and component concentrations are updated in each global iteration until a specified tolerance limit is reached. In each global iteration, phase and component transport equations are solved along streamlines and corrected for transverse fluxes on the underlying grid.

Our approach elegantly captures the effect of fluid and rock compressibility along streamlines as an extra source in each streamline segments. This improvement over previous streamline models helps to better improve on the pressure calculations resulting in superior equilibrium calculations. Furthermore, compared previous streamline models, formation dry out effects are better modelled in our approach. This is because, as recommended in this paper, mutual solubility computations were carried out along streamlines unlike in previous streamline-based methods. This enables more accurate prediction of well injectivity variations during CO2 injection especially for highly water wet media. These are evident in the series of examples (spanning all levels of dimensionality) presented in this paper where the results of our streamline-based simulation compares excellently well with commercial simulator (ECLIPSE-CO2STORE) results.

It is clear, based on our observation from the cases presented in this paper, that the nonlinearities associated with the multiphase-multicomponent problem can be decently reduced using our approach, resulting in fewer amount of time steps required for modeling the CO2 injection period. We however submit that the computational advantage achievable using this approach may reduce as transverse fluxes such as gravity, capillarity and molecular diffusion become more dominant.

# 8 Nomenclature

| **Symbols** | | **Meaning** |
|---|---|---|
| $a$ | = | Activity |
| $B$ | = | Surface fluid density to subsurface fluid density ratio |
| $c_t$ | = | Total compressibility |
| $\chi$ | = | Bi-streamfunction |
| $F$ | = | Phase fractional flow (ratio of mobility to total mobility) |
| $f$ | = | Phase fraction flow including transverse phase fluxes |
| $\gamma$ | = | Activity coefficient |
| $g$ | = | Gravitational acceleration |
| **k** | = | Permeability tensor |
| $\lambda$ | = | Fluid mobility |
| $\mu$ | = | Fluid viscosity |
| $m$ | = | mass |
| $M$ | = | Component mass per bulk volume |
| $MW$ | = | Molecular Weight |
| $\Phi$ | = | Fugacity coefficient |
| $\phi$ | = | Porosity |
| $\psi$ | = | Bi-streamfunction |
| $p$ | = | Pressure |
| $\rho$ | = | Fluid phase density |
| $\sigma$ | = | Flow divergence along streamlines |
| $s$ | = | Fluid phase saturation |
| $\tau$ | = | Convective Time of Flight (TOF) |
| $\mathbf{u}$ | = | Velocity field |
| $V_G$ | = | Grid volume |
| $x, \tilde{x}$ | = | Mass fraction & mole fraction in aqueous phase |
| $y, \tilde{y}$ | = | Mass fraction & mole fraction in CO2 phase |
| **Subscripts** | | **Meaning** |
| $aq$ | = | Aqueous phase |
| $Br$ | = | Brine |
| $c$ | = | CO2 component |
| $cd$ | = | Dissolved CO2 |
| $c\pi$ | = | Capillary pressure in phase $\pi$ |
| $g$ | = | CO2 gas phase |
| $\perp$ | = | Transverse direction |
| $s$ | = | Salt component |
| $w$ | = | Water component |